\documentclass[%
reprint,
superscriptaddress,
 amsmath,amssymb,
 aps,
prc,
altaffilletter
]{revtex4-1}

\usepackage{graphicx}
\usepackage{dcolumn}
\usepackage{bm}
\usepackage{hyperref}

\usepackage[dvipsnames]{xcolor} 
\usepackage{mathtools}

\usepackage{longtable}

\begin{document}

\preprint{APS/123-QED}

\title{Decay Spectroscopy of $^{129}$Cd}

\author{Y. Saito}
	\email{saito@triumf.ca}
	\affiliation{Department of Physics and Astronomy, University of British Columbia, Vancouver, British Columbia V6T 1Z4, Canada}
	\affiliation{TRIUMF, 4004 Wesbrook Mall, Vancouver, British Columbia V6T 2A3, Canada}
\author{I. Dillmann}
	\affiliation{TRIUMF, 4004 Wesbrook Mall, Vancouver, British Columbia V6T 2A3, Canada}
	\affiliation{Department of Physics and Astronomy, University of Victoria, Victoria, British Columbia V8P 5C2, Canada}
\author{R. Kr\"{u}cken}
	\affiliation{Department of Physics and Astronomy, University of British Columbia, Vancouver, British Columbia V6T 1Z4, Canada}
	\affiliation{TRIUMF, 4004 Wesbrook Mall, Vancouver, British Columbia V6T 2A3, Canada}
\author{N. Bernier}
	\affiliation{Department of Physics and Astronomy, University of British Columbia, Vancouver, British Columbia V6T 1Z4, Canada}
	\affiliation{TRIUMF, 4004 Wesbrook Mall, Vancouver, British Columbia V6T 2A3, Canada}
\author{G. C. Ball}
	\affiliation{TRIUMF, 4004 Wesbrook Mall, Vancouver, British Columbia V6T 2A3, Canada}	
\author{M. Bowry}
	\altaffiliation[Present address: ]{School of Computing, Engineering and Physical Sciences, University of the West of Scotland, Paisley PA1 2BE, Scotland, United Kingdom}
	\affiliation{TRIUMF, 4004 Wesbrook Mall, Vancouver, British Columbia V6T 2A3, Canada}	
\author{C. Andreoiu}
	\affiliation{Department of Chemistry, Simon Fraser University, Burnaby, British Columbia V5A 1S6, Canada}
\author{H. Bidaman}
	\affiliation{Department of Physics, University of Guelph, Guelph, Ontario N1G 2W1, Canada}
\author{V. Bildstein}
	\affiliation{Department of Physics, University of Guelph, Guelph, Ontario N1G 2W1, Canada}
\author{P. Boubel}
	\affiliation{Department of Physics, University of Guelph, Guelph, Ontario N1G 2W1, Canada}
\author{C. Burbadge}
	\affiliation{Department of Physics, University of Guelph, Guelph, Ontario N1G 2W1, Canada}
\author{R. Caballero-Folch}
	\affiliation{TRIUMF, 4004 Wesbrook Mall, Vancouver, British Columbia V6T 2A3, Canada}
\author{M. R. Dunlop}
	\affiliation{Department of Physics, University of Guelph, Guelph, Ontario N1G 2W1, Canada}
\author{R. Dunlop}
	\affiliation{Department of Physics, University of Guelph, Guelph, Ontario N1G 2W1, Canada}
\author{L. J. Evitts}
	\altaffiliation[Present address: ]{Nuclear Futures Institute, Bangor University, Bangor, Gwynedd, LL57 2DG, United Kingdom}
	\affiliation{TRIUMF, 4004 Wesbrook Mall, Vancouver, British Columbia V6T 2A3, Canada}
	\affiliation{Department of Physics, University of Surrey, Guildford, Surrey, GU2 7XH, United Kingdom}
\author{F. H. Garcia}
	\affiliation{Department of Chemistry, Simon Fraser University, Burnaby, British Columbia V5A 1S6, Canada}
\author{A. B. Garnsworthy}
	\affiliation{TRIUMF, 4004 Wesbrook Mall, Vancouver, British Columbia V6T 2A3, Canada}
\author{P. E. Garrett}
	\affiliation{Department of Physics, University of Guelph, Guelph, Ontario N1G 2W1, Canada}
\author{H. Grawe}
    \affiliation{GSI Helmholtzzentrum f\"{u}r Schwerionenforschung GmbH, 64291 Darmstadt, Germany}
\author{G. Hackman}
	\affiliation{TRIUMF, 4004 Wesbrook Mall, Vancouver, British Columbia V6T 2A3, Canada}
\author{S. Hallam}
	\affiliation{TRIUMF, 4004 Wesbrook Mall, Vancouver, British Columbia V6T 2A3, Canada}
	\affiliation{Department of Physics, University of Surrey, Guildford, Surrey, GU2 7XH, United Kingdom}
\author{J. Henderson}
	\altaffiliation[Present address: ]{Lawrence Livermore National Laboratory, Livermore, California 94550, USA}
	\affiliation{TRIUMF, 4004 Wesbrook Mall, Vancouver, British Columbia V6T 2A3, Canada}
\author{S. Ilyushkin}
	\affiliation{Department of Physics, Colorado School of Mines, Golden, Colorado 80401, USA}
\author{A. Jungclaus}
	\affiliation{Instituto de Estructura de la Materia, CSIC, E-28006 Madrid, Spain}
\author{D. Kisliuk}
	\affiliation{Department of Physics, University of Guelph, Guelph, Ontario N1G 2W1, Canada}
\author{J. Lassen}
	\affiliation{TRIUMF, 4004 Wesbrook Mall, Vancouver, British Columbia V6T 2A3, Canada}
	\affiliation{Department of Physics and Astronomy, University of Manitoba, Winnipeg, Manitoba R3T 2N2, Canada}
\author{R. Li}
	\affiliation{TRIUMF, 4004 Wesbrook Mall, Vancouver, British Columbia V6T 2A3, Canada}
\author{E. MacConnachie}
	\affiliation{TRIUMF, 4004 Wesbrook Mall, Vancouver, British Columbia V6T 2A3, Canada}
\author{A. D. MacLean}
	\affiliation{Department of Physics, University of Guelph, Guelph, Ontario N1G 2W1, Canada}
\author{E. McGee}
	\affiliation{Department of Physics, University of Guelph, Guelph, Ontario N1G 2W1, Canada}
\author{M. Moukaddam}
	\altaffiliation[Present address: ]{IPHC, CNRS/IN2P3, Universit\'{e} de Strasbourg, F-67037 Strasbourg, France}
	\affiliation{TRIUMF, 4004 Wesbrook Mall, Vancouver, British Columbia V6T 2A3, Canada}
\author{B. Olaizola}
	\altaffiliation[Present address: ]{TRIUMF, 4004 Wesbrook Mall, Vancouver, British Columbia V6T 2A3, Canada}
	\affiliation{Department of Physics, University of Guelph, Guelph, Ontario N1G 2W1, Canada}
\author{E. Padilla-Rodal}
	\affiliation{Universidad Nacional Aut\'{o}noma de M\'{e}xico, Instituto de Ciencias Nucleares, AP 70-543, M\'{e}xico City 04510, DF, M\'{e}xico}
\author{J. Park}
	\altaffiliation[Present Address: ]{Department of Physics, Lund University, 22100 Lund, Sweden}
	\affiliation{Department of Physics and Astronomy, University of British Columbia, Vancouver, British Columbia V6T 1Z4, Canada}
	\affiliation{TRIUMF, 4004 Wesbrook Mall, Vancouver, British Columbia V6T 2A3, Canada}
\author{O. Paetkau}
	\affiliation{TRIUMF, 4004 Wesbrook Mall, Vancouver, British Columbia V6T 2A3, Canada}
\author{C. M. Petrache}
	\affiliation{Centre de Sciences Nucl\'{e}aires et Sciences de la Mati\`{e}re, CNRS/IN2P3, Universit\'{e} Paris-Saclay, 91405 Orsay, France}
\author{J. L. Pore}
	\altaffiliation[Present address: ]{Lawrence Berkeley National Laboratory, Berkeley, CA 94720, USA}
	\affiliation{Department of Chemistry, Simon Fraser University, Burnaby, British Columbia V5A 1S6, Canada}
\author{A. J. Radich}
	\affiliation{Department of Physics, University of Guelph, Guelph, Ontario N1G 2W1, Canada}
\author{P. Ruotsalainen}
	\altaffiliation[Present address: ]{Department of Physics, University of Jyv\"{a}skyl\"{a}, P.O. Box 35, FI-40014, Jyv\"{a}skyl\"{a}, Finland}
	\affiliation{TRIUMF, 4004 Wesbrook Mall, Vancouver, British Columbia V6T 2A3, Canada}
\author{J. Smallcombe}
	\altaffiliation[Present address: ]{Oliver Lodge Laboratory, The University of Liverpool, Liverpool, L69 7ZE, United Kingdom}
	\affiliation{TRIUMF, 4004 Wesbrook Mall, Vancouver, British Columbia V6T 2A3, Canada}
\author{J. K. Smith}
	\altaffiliation[Present address: ]{Division of Natural Sciences, Pierce College Puyallup, Puyallup, WA 98374, USA}
	\affiliation{TRIUMF, 4004 Wesbrook Mall, Vancouver, British Columbia V6T 2A3, Canada}
\author{C. E. Svensson}
	\affiliation{Department of Physics, University of Guelph, Guelph, Ontario N1G 2W1, Canada}	
\author{A. Teigelh\"{o}fer}
	\affiliation{TRIUMF, 4004 Wesbrook Mall, Vancouver, British Columbia V6T 2A3, Canada}
	\affiliation{Department of Physics and Astronomy, University of Manitoba, Winnipeg, Manitoba R3T 2N2, Canada}
\author{J. Turko}
	\affiliation{Department of Physics, University of Guelph, Guelph, Ontario N1G 2W1, Canada}
\author{T. Zidar}
	\affiliation{Department of Physics, University of Guelph, Guelph, Ontario N1G 2W1, Canada}
	
\date{\today}

\begin{abstract}
Excited states of $^{129}$In populated following the $\beta$-decay of $^{129}$Cd were experimentally studied with the GRIFFIN spectrometer at the ISAC facility of TRIUMF, Canada. A 480-MeV proton beam was impinged on a uranium carbide target and $^{129}$Cd was extracted using the Ion Guide Laser Ion Source (IG-LIS). $\beta$- and $\gamma$-rays following the decay of $^{129}$Cd were detected with the GRIFFIN spectrometer comprising the plastic scintillator SCEPTAR and 16 high-purity germanium (HPGe) clover-type detectors. 
From the $\beta$-$\gamma$-$\gamma$ coincidence analysis, 32 new transitions and 7 new excited states were established, expanding the previously known level scheme of $^{129}$In. The $\log ft$ values deduced from the $\beta$-feeding intensities suggest that some of the high-lying states were populated by the $\nu 0 g_{7/2} \rightarrow \pi 0 g_{9/2}$ allowed Gamow-Teller (GT) transition, which indicates that the allowed GT transition is more dominant in the $^{129}$Cd decay than previously reported. Observation of fragmented Gamow-Teller strengths is consistent with theoretical calculations.

\end{abstract}

\maketitle


\section{\label{sec:Intro}Introduction}
 The nuclei around doubly magic $^{132}$Sn have been the topic of significant recent interest in order to study the validity and limits of the nuclear shell model and realistic nuclear forces \cite{nature,laser132Sn,Kozub2012,Sarkar2001,Kartamyshev2007,Coraggio2009}.
 Among experimentally accessible nuclei, $^{132}$Sn is the heaviest unstable nucleus with both proton and neutron closed shells.  Furthermore, the $N=82$ neutron shell closure is responsible for the second abundance peak of the astrophysical rapid neutron capture process ($r$-process) at $A\approx130$ \cite{synthesis}. Recent $r$-process sensitivity studies have shown that experimental information on nuclei near $N=82$ has significant impact on $r$-process abundance calculations \cite{mumpower2016}. 

Two low-lying $\beta$-decaying states in $\prescript{129}{48}{\mathrm{Cd}}_{81}^{\,}$ have previously been observed, and their spin and parities have been identified as $11/2^-$ and  $3/2^+$. These spin assignments were confirmed through laser spectroscopy \cite{yordanov}. Refs.~\cite{kratz2005, wangshellmodel} suggested $11/2^-$ as the ground state and an excited $3/2^+$ $\beta$-decaying isomer, and this ordering was confirmed by a mass measurement with the ISOLTRAP spectrometer at ISOLDE-CERN, placing the $3/2^+$ state at 343(8)~keV \cite{ISOLTRAP}.

The half-lives of these two $\beta$-decaying states in $^{129}$Cd have been measured in three different experiments. \citeauthor{arndt} \cite{arndt} reported half-lives of 104(6) ms for the $11/2^-$ state and 242(8) ms for the $3/2^+$ state, respectively, by measuring $\beta$-delayed neutrons at CERN-ISOLDE. Measurements utilizing the EURICA setup at the Radioactive Isotope Beam Factory (RIBF) at RIKEN Nishina Center (Japan) by \citeauthor{EURICA} reported values of 155(3) ms for the $11/2^-$ and 148(8) ms for the $3/2^+$ states, respectively, which disagree strongly with the previous measurement \cite{EURICA}. Finally, using the same dataset recorded with the GRIFFIN spectrometer as presented here, \citeauthor{CdHalflife} \cite{CdHalflife} confirmed the EURICA results with half-lives of 147(3) ms for the $11/2^-$ state and 157(8) ms for the $3/2^+$ state. The two recent measurements reported in Refs.\cite{EURICA,CdHalflife} both used the $\beta$-$\gamma$-gating method and are in agreement within $2\sigma$ for the $11/2^-$ and within $1\sigma$ for the $3/2^+$ state, respectively.

The previous studies also included $\gamma$-ray spectroscopy of the decay of $^{129}$Cd. \citeauthor{arndt} identified more than 50 $\gamma$-ray transitions following the $\beta$-decay of $^{129}$Cd, confirming the placement of the $17/2^-$ isomeric state with a half-life of 8.5(5) $\mu$s at $1687$ keV reported by  \citeauthor{isomer} \cite{isomer} and the $\beta$-decaying $1/2^-$ isomer reported in Refs. \cite{geer, decay129In, kankainen}. This $1/2^-$ isomeric state was also confirmed by a recent mass measurement at the TITAN facility at TRIUMF-ISAC \cite{titan}, reporting an excitation energy of 444(15) keV, in agreement with the previous result \cite{kankainen}. 

\citeauthor{EURICA} recently expanded the level scheme of $^{129}$In with 32 newly observed transitions, resulting in establishing 27 new excited states  \cite{EURICA}. In addition, $\beta$-feeding intensities and $\log{ft}$ values were reported for the first time. The experimental results were compared to shell-model calculations for excited states below 1.5 MeV.

In the current study, we aim to resolve the discrepancies in the level scheme of $^{129}$In reported by the previous studies \citep{arndt,EURICA} with the high statistics of the current GRIFFIN data set. The properties of the $\beta$-decay of $^{129}$Cd are also studied in detail, and the deduced Gamow-Teller strengths $B(\mathrm{GT})$ are compared to theoretical calculations.

\section{\label{sec:ExpMet}Experimental Method}
The decay of $^{129}$Cd was studied with the GRIFFIN spectrometer \cite{GRIFFINarray, Svensson, HPGe} at the TRIUMF-ISAC facility \cite{Dilling}. 
The $^{129}$Cd ions were produced using a 480-MeV proton beam from the TRIUMF main cyclotron impinged on a UC$_{x}$ target. The Ion-Guide Laser Ion Source (IG-LIS) \cite{IGLIS} was used for the extraction of the isotopes of interest. IG-LIS suppresses surface-ionized species such as In and Cs with repeller electrodes after the initial production and diffusion of isotopes from the target. Neutral $^{129}$Cd atoms can pass the repeller electrodes and are then ionized by element-selective multi-step laser excitation. The $^{129}$Cd ions were then extracted, accelerated to 28 keV, and selected by a high-resolution mass separator. The beam of $^{129}$Cd, which consists of both, the ground state and the $\beta$-decaying isomeric state, was delivered to the GRIFFIN spectrometer with an intensity of $\approx$120 pps for approximately 13 hours. 

\begin{figure}
\includegraphics[width=\linewidth]{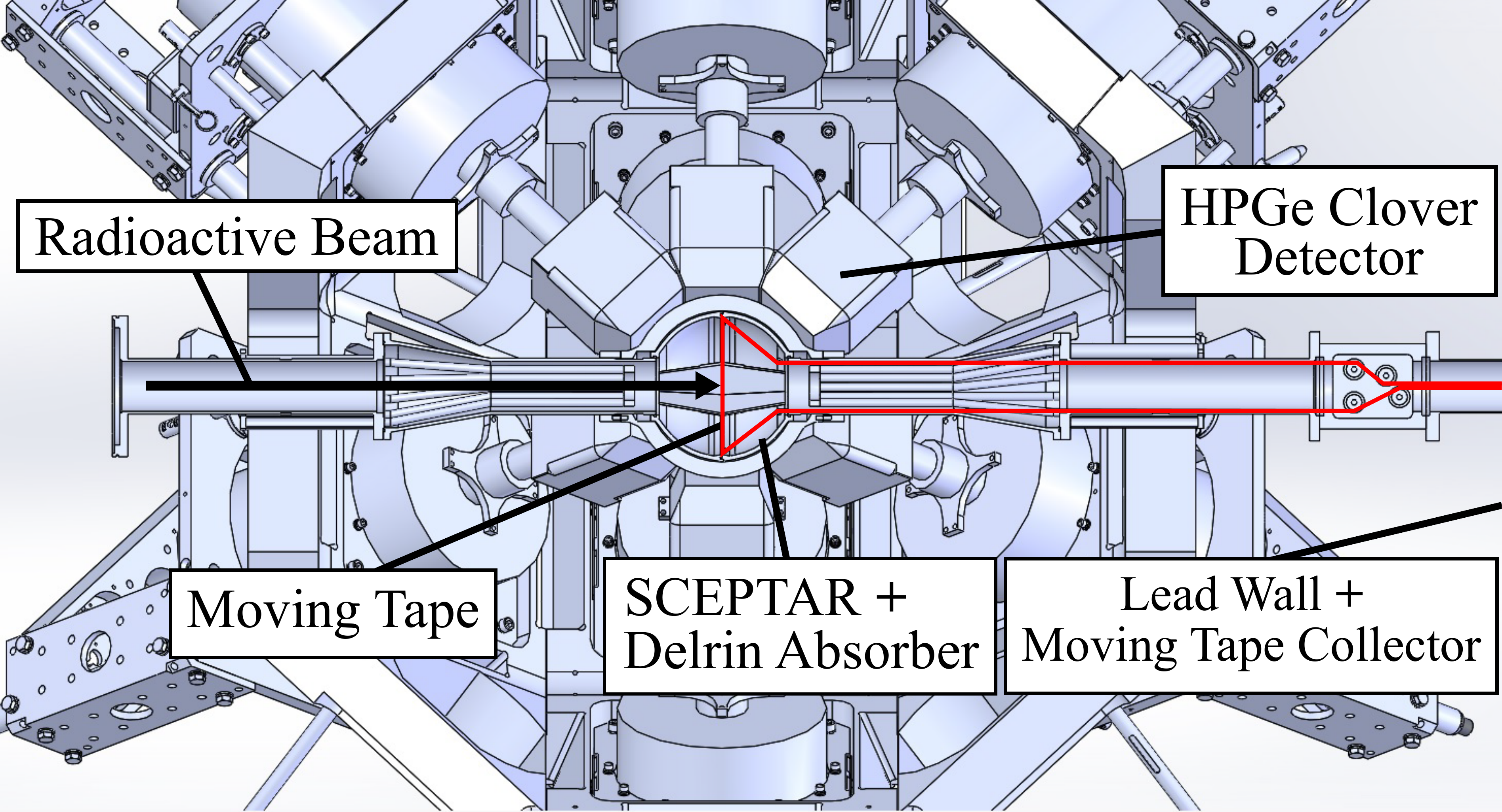}
\caption{\label{fig:GRIFFINsetup}{The radioactive ion beam delivered to the spectrometer is implanted at the mutual centers of the detector arrays into the tape which is then moved and collected behind a lead wall (not shown in the figure). }}
\end{figure}

The GRIFFIN array consists 16 large-volume high-purity germanium (HPGe) clover detectors \cite{HPGe}. SCEPTAR, an array of 20 thin plastic scintillators, was placed at the center of GRIFFIN in order to detect $\beta$-particles and thus be able to tag $\gamma$-rays following the decay of $^{129}$Cd \cite{GRIFFINarray}. SCEPTAR is surrounded by a Delrin absorber with a thickness of 20 mm for the suppression of bremsstrahlung. The beam was implanted into a movable aluminized mylar tape, which is a component of the moving tape collector, located at the mutual center of GRIFFIN and SCEPTAR (FIG. \ref{fig:GRIFFINsetup}). The maximum $\beta$-$\gamma$ correlation time was set to 800~ns, and by using $\beta$-$\gamma$ coincidences, background $\gamma$-rays which are not associated with $\beta$-decays within the array can be suppressed.

The data were collected in a cycle mode, which consists of a background measurement, a beam implantation period, where the ions in the beam are collected onto the tape, a decay period with the beam blocked by the electrostatic kicker upstream of GRIFFIN, and a tape move period, where the tape with the unwanted decay products are moved to a shielded position outside of the array.

The absolute photopeak detection efficiencies of the GRIFFIN array for a 1-MeV and 4-MeV $\gamma$-ray with the 20~mm Delrin absorber in 11~cm distance were determined to be 9.4(3)\% and 3.3(2)\%, respectively.

\section{\label{sec:ExpRes}Experimental Results}
\begin{figure}
\includegraphics[width=\linewidth]{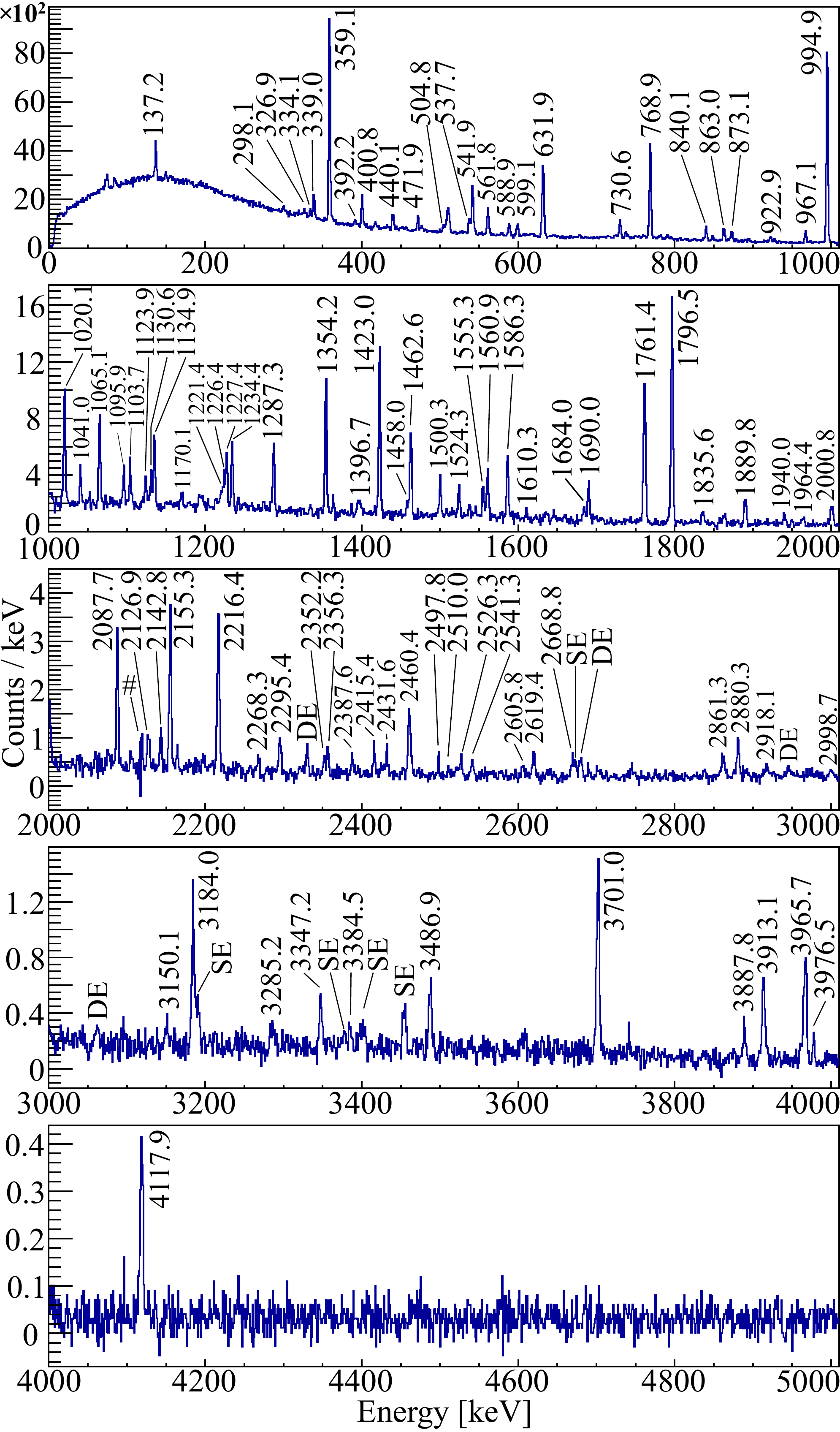}
\caption{\label{fig:bgsingles}(Color online) The $\beta$-gated $\gamma$-ray spectrum with $^{129}$Sn transitions subtracted (see text for details). Single-escape peaks and double-escape peaks are labeled with ``SE'' and ``DE'', respectively. The ``\#'' symbol denotes the residue of the subtraction from a strong 2118.2~keV transition in $^{129}$Sn.}
\end{figure}

In order to identify transitions belonging to the $\beta$-decay of $^{129}$Cd to $^{129}$In, $\gamma$-rays following the decay of the daughter nucleus $^{129}$In to $^{129}$Sn were subtracted from the $\gamma$-ray spectrum. For the experimental runs a data collection cycle consisting of 0.5~s of background measurement, 0.6~s of beam implantation (beam-on), and 1.5~s of decay (beam-off) was used. Due to the shorter $\beta$-decay half-lives for the ground and isomeric state of $^{129}$Cd ($\approx$150~ms) \cite{CdHalflife, EURICA} compared to the $\beta$-decay half-lives of $^{129}$In of 611~ms ($9/2^{+}$ ground state) and 1.23~s ($1/2^-$ $\beta$-decaying isomeric state) \cite{ENSDF}, the number of the decays of $^{129}$Cd are dominating in the beam-on period, and then $^{129}$In becomes dominant for most of the 1.5~s decay (beam-off) period. Therefore, by subtracting the $\gamma$-ray spectrum from the ``beam-off'' from the overall $\gamma$-ray spectrum with appropriate scaling, a $\gamma$-ray spectrum free of transitions belonging to the granddaughter nucleus $^{129}$Sn can be obtained. The width of the ``beam-off'' time window is determined so that the ratio of the number of decays of the $1/2^-$ state and the $9/2^+$ state in the time window becomes the same as the corresponding ratio in the whole spectrum to account for the time dependence of the activities of the two states in $^{129}$In. The uncertainty arising from the scaling and the subtraction is propagated to each bin content of the spectrum. The resulting $\gamma$-ray spectrum is shown in FIG. \ref{fig:bgsingles}. 

The maximum $\beta$-$\gamma$ correlation time is chosen to be 800 ns so that some of the 334.1 keV isomeric transition can be captured in the $\beta$-gated $\gamma$-ray spectrum and its contribution to the intensity of the 359.1, 994.9, and 1354.2 keV transitions can be subtracted. This procedure is necessary to scale the relative intensities of 334.1, 359.1, 994.9, and 1354.2 keV transitions in order to obtain the number of those $\gamma$-rays in the $\beta$-gated $\gamma$-ray spectrum with an internal conversion (IC) correction.

\subsection{\label{subsec:levels}Level scheme of $^{129}$In}
\begin{figure*}
\includegraphics[width=\linewidth]{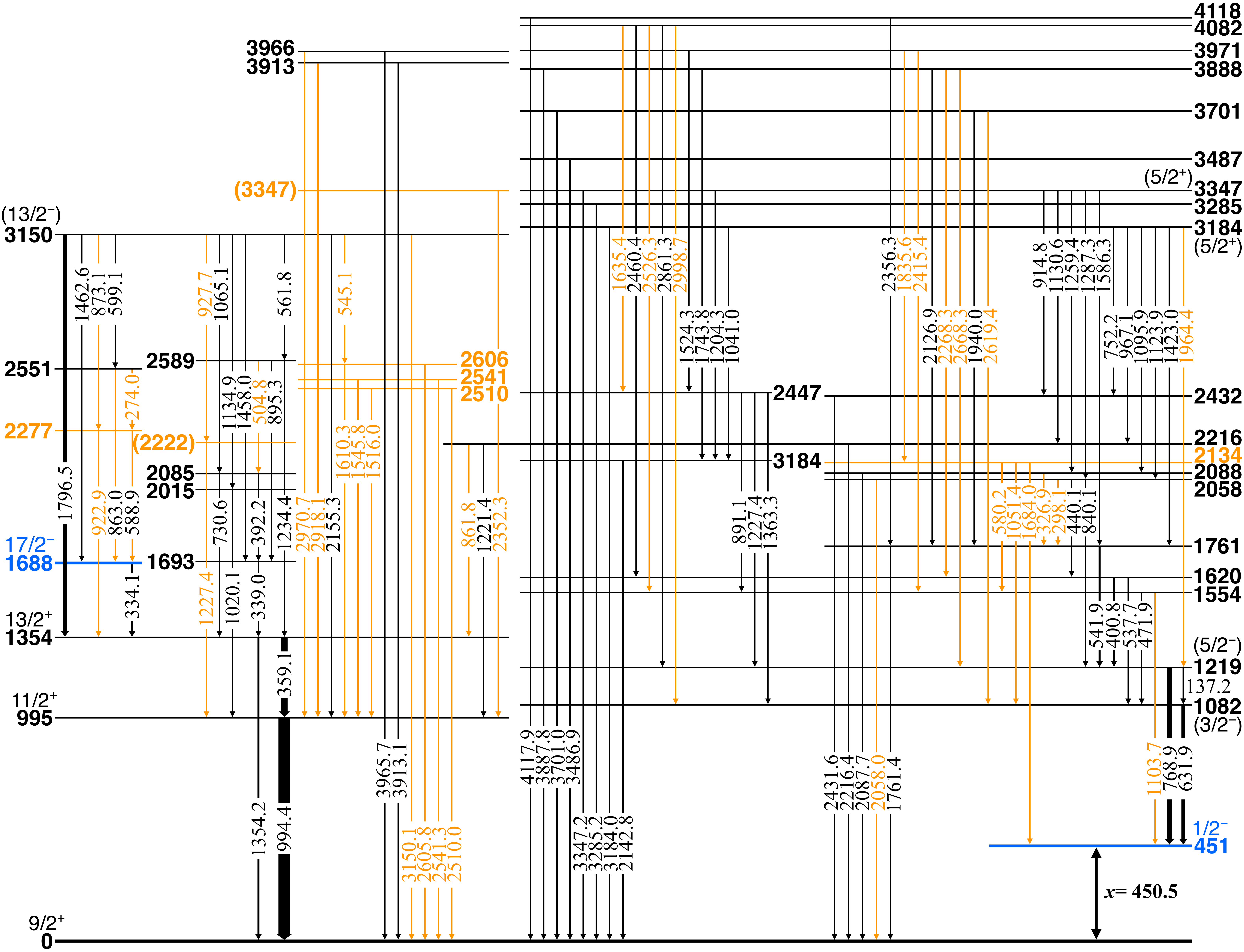}
\caption{\label{fig:levelscheme}(Color online) The proposed level scheme of $^{129}$In. Newly identified states and transitions are indicated by light grey (orange) color. Dark grey (blue) horizontal lines for the states at 451 and 1688 keV indicate isomeric states. Previously observed states or transitions are indicated by black color.}
\end{figure*}

\begin{figure}
\includegraphics[width=\linewidth]{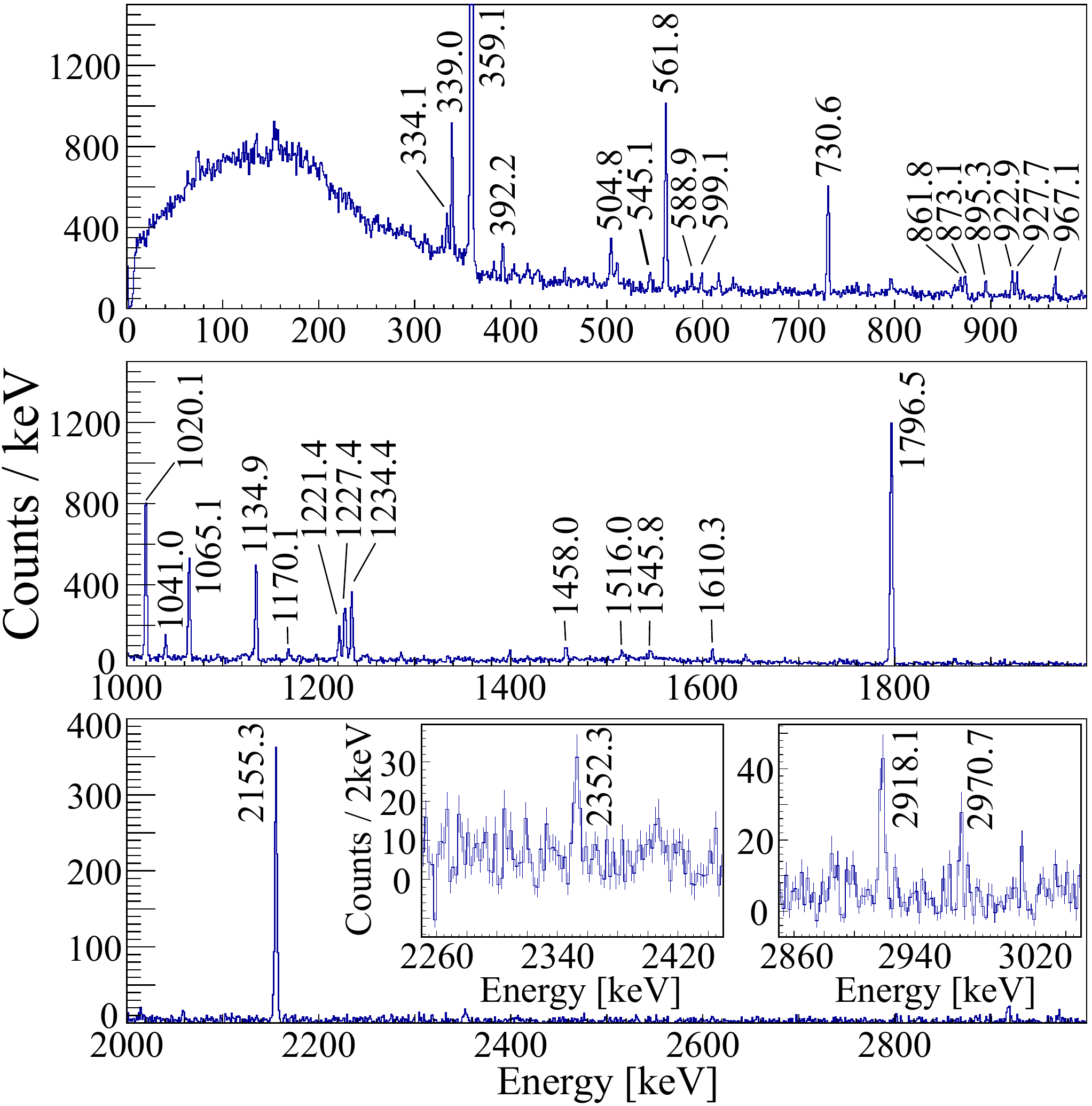}
\caption{\label{fig:995gate}(Color online) Observed $\gamma$-ray spectrum in coincidence with $\beta$-particles and the 994.9 keV transition. 
}
\end{figure}

As a result of the analysis of the $\beta$-$\gamma$ coincidence spectra as well as the $\beta$-$\gamma$-$\gamma$ coincidence spectra, 118 $\gamma$-ray transitions, including 32 new ones were identified and placed in the level scheme of $^{129}$In. Five excited states were newly established. The level scheme is shown in FIG. \ref{fig:levelscheme}. Starting from the most intense 994.9 keV transition, the level scheme was constructed based on the $\gamma$-$\gamma$ coincidence information, as shown in FIG. \ref{fig:995gate}. 

In general, the level scheme proposed in this work is in agreement with \citeauthor{EURICA} \citep{EURICA}. In the subsequent sections, we will only discuss the placement of transitions that differ from the previously reported results, as well as new transitions and excited states.

\subsubsection{Excited states feeding the $13/2^+$ and $17/2^-$ levels \\ at 1354 and 1688 keV}
While the placement of the $17/2^-$ isomer at 1688 keV in $^{129}$In was previously confirmed \citep{isomer, arndt, EURICA}, the placement of transitions feeding the $17/2^-$ isomer are in agreement except for the 873 keV (or 588.9 keV) transition feeding the isomer \citep{arndt, EURICA}. \citeauthor{EURICA} \citep{EURICA} tentatively established excited states at 2551 keV and 2561 keV, placing the 863.0 keV and the 873.1 keV transitions on top of the $17/2^-$ state. These transitions were proposed to be in prompt coincidence with the 599.1 and 588.9 keV transitions, respectively. However, the order of these $599.1 + 863.0$ keV and $588.9 + 873.1$ keV cascades could not be determined in previous experiments.

In the current experiment, a 922.9 keV transition is observed in coincidence with the 873.1 keV transition. The energy sum of the 334.1(5) and 588.9(5) keV transitions matches the 922.9(5) keV transition within uncertainties. Since the 334.1 keV isomeric transition is not in prompt coincidence with the 588.9 keV transition, the 922.9 keV is not a result of the summing of the two transitions. Therefore, the 922.9 keV transition is placed on top of the $13/2^+$ state, bypassing the $17/2^-$ isomer, establishing a new excited state at 2277 keV and confirming the 2551 keV level. This placement is further supported by the newly observed 274.0 keV transition in coincidence with 588.9 and 599.1 keV transitions. Since the 274.0 keV transition is not observed to be in coincidence with the 863.0 keV transition, it is proposed to be in parallel with the transition. Therefore, the 599.1 keV state is placed to feed the 2551 keV level. 
 
\subsubsection{Excited states feeding the $11/2^+$ level at 995 keV \\ and the $9/2^+$ ground state}
Ground-state transitions with energies of 3965.7 and 3913.1 keV were previously identified \citep{EURICA} and are confirmed in the current study. In addition, in this experiment, 2970.7 and 2918.1 keV transitions are newly observed in coincidence with the 994.9 keV ground state transition. The energy sum of these 2970.7(10) and 2918.1(10) keV transitions with the 994.9(5) keV transitions are 3965.7(10) and 3913.1(10) keV, respectively. Therefore, we place these transitions in parallel to the 3965.7 and 3913.1 keV transitions, which further supports the placement of these transitions as ground-state transitions.

Similarly, we observe new transitions with the energies of 1610.3(5), 1545.8(6), and 1516.0(6) keV in coincidence with the 994.9(5) keV transition. The energy sum of these cascades are 2605.8(6), 2541.3(6), and 2510.0(6) keV, respectively. Transitions corresponding to these energy sums are also observed at 2605.8(6), 2541.3(6), and 2510.0(6) keV. Therefore, we propose new excited states at 2606, 2541, and 2510 keV. The 2605.8, 2541.3, and 2510.0 keV transitions are placed in the level scheme as ground-state transitions. The 1610.3, 1545.8, and 1516.0 keV transitions are placed to feed the $11/2^+$ state at 995 keV, in parallel to those ground-state transitions.

As shown in Fig. \ref{fig:995gate}, a 1227.4 keV transition is observed in coincidence with the 994.9 keV transition. This 1227.4 keV $\gamma$-ray is also in coincidence with the 927.7 keV transition. The energy sum of the 994.9, 1227.4, and 927.7 transitions is 3150 keV, therefore it is highly likely that the 1227.4 and 927.7 keV transitions follow the de-excitation of the ($13/2^-$) state at 3150 keV. The order of these transitions is tentatively determined based on their relative transition intensities, which are 1.0(4) \% for the 927.7 keV transition and 3.9(17) \% for the 1227.4 keV transition, resulting in tentatively establishing a level at 2222 keV.

\subsubsection{The $1/2^-$ $\beta$-decaying isomer at 451 keV}
The placement of the 631.9 and 768.9 keV transitions on top of the $1/2^-$ isomer was previously proposed in Ref.~\citep{EURICA} and is confirmed in the current experiment based on the $\gamma$-$\gamma$ coincidence information and the relative intensities. In addition to the 631.9 and 768.9 keV transition, we observe two other transitions with energies of 1103.7 and 1684.0 keV that feed this isomer. The placement of the 1103.7 keV transition is confirmed by its coincidence with the 891.1 keV transition. This is consistent with the energy difference between the 1554 keV level and the $1/2^-$ isomer at 451 keV. The 1684.0 keV transition is discussed in the following section.  

The (exact) energy of the $1/2^-$ $\beta$-decaying isomer can be determined using the transitions following the de-excitation of the 1761 keV state, since the 1761.4 keV transition feeds the $9/2^+$ ground state and the 768.9 keV transition following the 541.9 keV transition feeds the $1/2^-$ isomer. The excitation energy is deduced to be 450.5(8) keV. This value is in perfect agreement with the previously reported values of 451 (1) keV based on $\gamma$-ray spectroscopy \citep{EURICA} and 444 (15) keV from the mass measurement by the TITAN collaboration \citep{titan}. However, another mass measurement by JYFLTRAP reported a slightly higher value of 459 (5) keV \citep{kankainen}.

\subsubsection{The 2134 keV level}
\begin{figure}
\includegraphics[width=\linewidth]{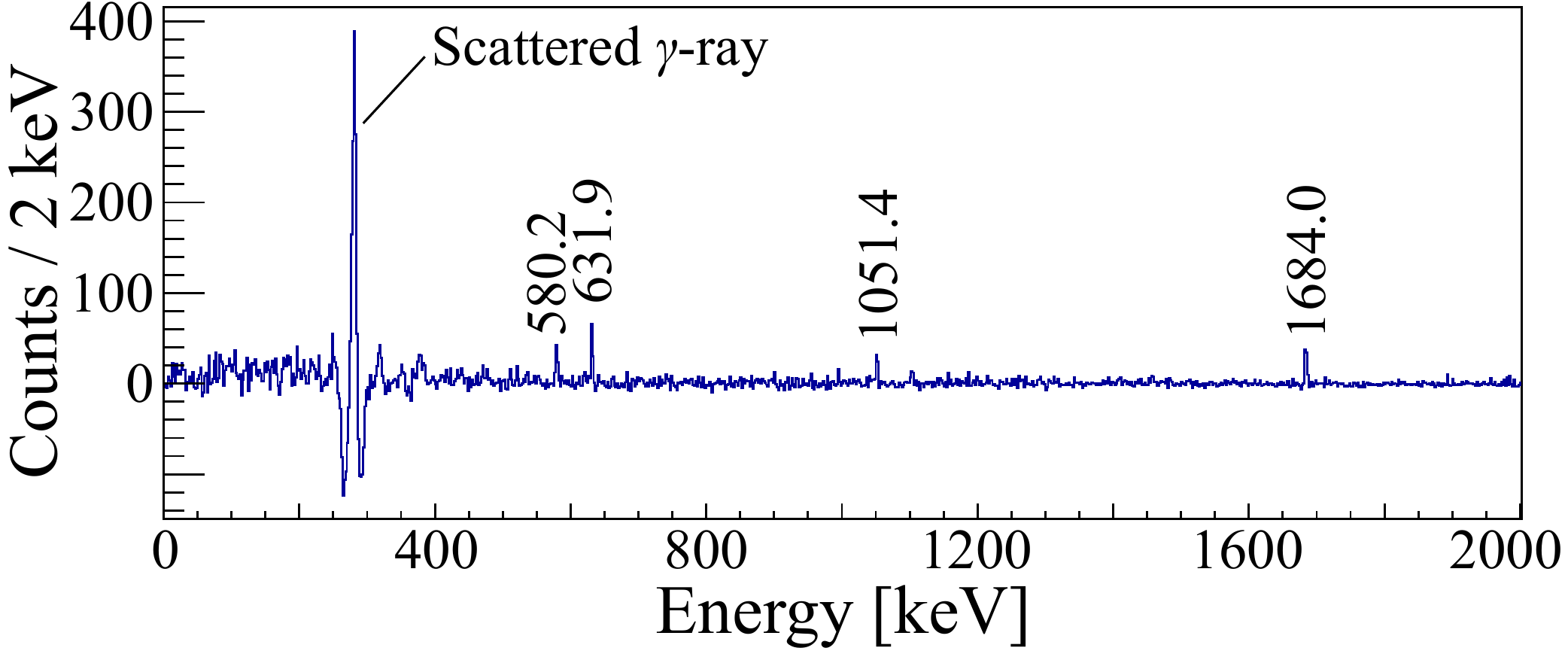}
\caption{\label{fig:1835gate}(Color online) Observed $\gamma$-ray spectrum in coincidence with $\beta$-particles and the 1835.6 keV transition, which provides evidence of the newly established level at 2134 keV. 
The ``scattered $\gamma$-ray'' peak arises from the strong 2118.3 keV transition \citep{ENSDF} in $^{129}$Sn, when part of the energy is absorbed in one HPGe crystal and the rest is detected by another HPGe crystal.}
\end{figure}

A new transition with an energy of 1051.4 keV was observed in coincidence with the 631.9 keV transition. We do not see this transition in coincidence with the 768.9 keV transition, which implies that the 1051.4 keV transition feeds the 1082 keV level. This 1051.4 keV $\gamma$-ray is observed to be in coincidence with the 1835.6 keV transition as well. The spectrum gated on this 1835.6 keV transition shows peaks at 580.2, 631.9, 1051.4, and 1684.0 keV, as shown in FIG. \ref{fig:1835gate}. 

Since the energy difference between 1684.0(5) and 1051.4(5) keV is 632.6(7) keV, the 1684.0 keV transition directly feeds the $1/2^-$ isomer at 451 keV, which is consistent with our placement of the 1051.4 keV transition. Furthermore, since the energy sum of the 1684.0(5) and 1835.6(5) keV transitions and the excitation energy of the isomer at 450.5(8) keV is 3970.1(11) keV, the 1835.6 keV transition is placed depopulating the 3971 keV level. The 580.2 keV $\gamma$-ray can be placed to feed the 1554 keV level, which is consistent with its observation in coincidence with the 471.9 keV transition.

\subsubsection{\label{subsubsec:2352keV}The 2352.3 keV transition}
The 3347 keV level has been proposed to be populated by the $\nu 0 g_{7/2} \rightarrow \pi 0 g_{9/2}$ allowed GT transition from the decay of $3/2^+$ state in $^{129}$Cd \cite{EURICA}. Our work supports this hypothesis (see Section \ref{sec:discussion}). Since a 3347.2 keV transition directly to the $9/2^+$ ground state is observed, its spin and parity were tentatively assigned to be ($5/2^+$). 

In the current experiment, a 2352.3(6) keV transition with a relative intensity of 0.7(2)~\% is observed in coincidence with the 994.9(5) keV transition (see FIG. \ref{fig:995gate}). The sum of the two energies is 3347.2(8) keV and matches well with the 3347.2(10) keV ground-state transition, therefore one possibility is to place the 2352.3 keV transition between the ($5/2^+$) state at 3347 keV state and the $11/2^+$ state at 995 keV. 

However, the lowest multipolarity of this transition would be $\mathrm{M3}$, which is expected to be highly hindered. Assuming that this transition is pure $\mathrm{M3}$ and the 3347.2-keV ground-state transition is pure $\mathrm{E2}$, the ratio of the reduced transition probability is deduced to be $B(\mathrm{M3})/B(\mathrm{E2})=7.5(15)\times 10^7\,[{\mu_N}^2/e^2]$, where $\mu_N$ is the nuclear magneton and $e$ is the elementary charge. This value is extremely large compared to the value calculated with the Weisskopf single-particle estimate, which is $B_\mathrm{sp}(\mathrm{M3})/B_\mathrm{sp}(\mathrm{E2})=27.8\,[{\mu_N}^2/e^2]$. This implies that this 2352.3 keV transition is not a $\mathrm{M3}$ transition and the experimental information points to the existence of a second state at 3347 keV which is populated by the decay of the $11/2^-$ state in $^{129}$Cd.

\subsection{\label{subsec:intensity}Relative $\gamma$-ray intensities}
For most of the $\gamma$-ray transitions, the relative intensities were determined by fitting the peaks in the $\beta$-gated $\gamma$-ray spectrum with appropriate background subtraction (FIG. \ref{fig:bgsingles}). The intensities of the weaker transitions were obtained from the $\beta$-gated $\gamma$-$\gamma$ coincidence spectra. 
The relative intensities of the 334.1, 359.1, 994.9, and 1354.2 keV transitions are determined from the $\gamma$-singles spectrum, which do not require any correlation with $\beta$ particles nor other $\gamma$-rays. The list of transitions and their relative intensities are shown in Table \ref{tab:gamma}.

\begingroup
\squeezetable
\begin{table*}
\caption{\label{tab:gamma} List of $\gamma$-ray transitions (in keV) and relative intensities (relative to the 994.9 keV transition). The values for $I_{\gamma}^{\text{lit}}$ are taken from Ref.~\cite{EURICA}. The uncertainties for the transition energies include the statistical uncertainty and the systematic uncertainty arising from energy calibration. Above the last energy calibration point (2614.51(1) keV, natural background $\gamma$-ray from the decay of $^{208}$Tl \cite{ENSDF}), the systematic uncertainty on the energy is evaluated to be up to 1 keV. Listed relative transition intensities are corrected for summing effects \cite{GRIFFINarray}. The column for $I_{\gamma + \text{ce}}$ shows the relative transition intensities corrected for internal conversion using the program BrIcc v2.3S \citep{bricc}. The corrections are applied for transitions below 1~MeV. When the spins and parities of both initial and final states are known, corrections are made up to electric/magnetic quadrupole ($\mathrm{E2}$/$\mathrm{M2}$) transitions. Otherwise, the conversion coefficient is assumed to be the average of E1 and M1 transitions, and a half of their differences are considered to be a systematic error. Observed transitions which are not placed in the level scheme in FIG.~\ref{fig:levelscheme} are marked by an asterisk (*).}
\begin{ruledtabular}
\begin{tabular}{lccccccc | lcccccc}

 $E_\gamma$ [keV] & $I_\gamma$ & $I_{\gamma + \text{ce}}$ & $I_\gamma^\text{lit}$ & $E_i$ [keV] & $E_f$ & $J_i^\pi$ & $J_f^\pi$ & $E_\gamma$ [keV] & $I_\gamma$ & $I_\gamma^\text{lit}$ & $E_i$ & $E_f$ & $J_i^\pi$ & $J_f^\pi$ \\ \hline
 137.2(5)		& 5.8(5)    & 7.9(10)
                                            & 4.7(6)    & 1220 	& 1082 	& $(5/2^-)$ 	& $(3/2^-)$	&1516.0(6) 		& 1.0(2) 	&  		    & 2510	& 995	& & $11/2^+$\\
 274.0(5)		& 1.1(2)    & {1.1(2)}
                                            &		    & 2551 	& 2277 	&			    &			&1524.3(5) 		& 2.8(5)	& 2.3(4)	& 3971	& 2447	& & \\ 
 298.1(5)		& 1.1(2)	& {1.1(2)}
                                            &			& 2058 	& 1761 	& 			    & 			&1537.9(5)$^*$ 	& 0.9(4)	& 		    & 		& 		& & \\ 
 326.9(5)		& 3.2(3)	& {3.3(3)}
                                            &		    & 2088 	& 1761 	& 			    & 			&1545.8(6) 		& 0.5(3) 	& 		    & 2541	& 995	& & $11/2^+$ \\ 
 334.1(5)		& 17.3(8)   & 18.7(8)
                                            & 19.9(12)	& 1688 	& 1354	& $17/2^-$	    & $13/2^+$	&1555.3(5)$^*$ 	& 3.0(5) 	& 1.3(4) 	& 		& 		& & \\ 
 339.0(5)		& 5.1(7)    & {5.2(7) }
                                            & 6.5(7) 	& 1693 	& 1354 	&			    & $13/2^+$	&1560.9(5)$^*$ 	& 5.5(6) 	& 2.7(11) 	& 		& 		& &\\ 
 359.1(5)		& 55.0(23)  & 56.0(23)	
                                            & 57.4(30)	& 1354 	& 995	& $13/2^+$	    & $11/2^+$	&1586.3(5) 		& 7.3(6) 	& 8.3(8) 	& 3347	& 1761	& $(5/2^+)$ & \\ 
 392.2(5)		& 1.7(3) 	& {1.7(3)}
                                            &1.4(7) 	& 2085	& 1693 	& 		    	& 			&1610.3(5) 		& 0.9(4) 	& 		    & 2606	& 995	& & $11/2^+$\\ 
 400.8(5)		& 7.8(4)    & {7.8(4)}
                                            & 6.3(6) 	& 1620	& 1220 	& 			    & $(5/2^-)$	&1635.4(5) 		& 1.7(8) 	& 		    & 4082	& 2447	& & \\ 
 440.1(5)		& 4.5(3) 	& {4.6(3)}
                                            & 2.7(3)    & 2058 	& 1620 	&    			& 			&1659.8(6)$^*$ 	& 0.8(5) 	&    		& 		& 		& &\\ 
 471.9(5)		& 4.5(3)    & {4.5(3)}
                                            & 1.5(8) 	& 1554 	& 1082 	&   			& $(3/2^-)$	&1684.0(5) 		& 2.1(5) 	&	    	& 2134	& 451	& & $1/2^-$ \\ 
 504.8(5)		& 2.3(4) 	& {2.3(4)}
                                            & 2.1(2)    & 2589 	& 2085 	& 	    		& 			&1690.0(5)$^*$ 	& 4.7(5) 	& 4.5(5) 	& 		& 		& & \\
 537.7(5)		& 4.6(3)    & {4.6(3)}
                                            & 2.8(8) 	& 1620 	& 1082 	& 		    	& $(3/2^-)$	&1743.8(6) 		& 0.4(1)	& 1.2(6) 	& 3888	& 2143	& & \\ 
 541.9(5)		& 16.4(8)   & {16.4(8) }
                                            & 14.5(10)	& 1761 	& 1220 	& 			    & $(5/2^-)$	&1761.4(5) 		& 16.7(10)	& 16.4(12)  & 1761	& 0		& & $9/2^+$\\
 545.1(5)		& 0.8(1)    & {0.8(1) }
                                            & 			& 3150	& 2606	& $(13/2^-)$	&			&1796.5(5) 		& 29.2(17)	& 26.4(15)	& 3150 	& 1354 	& $(13/2^-)$ & $13/2^+$ \\ 
 561.8(5)		& 9.5(7)    & {9.6(7) }
                                            & 8.5(8) 	& 3150 	& 2589 	& $(13/2^-)$	& 			&1835.6(5) 		& 1.6(4) 	& 		    & 3971	& 2134	& & \\ 
 580.2(5) 		& 1.0(5) 	& {1.0(5)}
                                            & 			& 2134 	& 1554 	& 			    & 			&1889.8(5)$^*$ 	& 3.4(5) 	& 5.0(6) 	& 		& 		& & \\ 
 588.9(5) 		& 4.5(5)    & {4.5(5) }
                                            & 3.2(4)    & 2277 	& 1688	& 			    & $17/2^-$	&1940.0(5) 		& 1.5(4) 	& 0.6(4) 	& 3701	& 1761	& & \\ 
 599.1(5) 		& 4.6(5)    & {4.6(5) }
                                            & 2.8(4) 	& 3150 	& 2551 	& $(13/2^-)$	& 			&1964.4(6) 		& 1.6(5) 	& 		    & 3184	& 1220	& $(5/2^+)$ & $(5/2^-)$ \\ 
 631.9(5) 		& 28.4(14)  & 28.5(14) 	    
                                            & 21.0(12)	& 1082 	& 451 	& $(3/2^-)$	    & $1/2^-$	&2000.8(5)$^*$	& 3.3(4)	& 5.3(12)	& 		& 		& & \\
 730.6(5) 		& 7.1(6)    & {7.1(6) }
                                            & 6.2(7)	& 2085 	& 1354 	& 		    	& $13/2^+$	&2058.0(6) 		& 0.6(2) 	& 		    & 2058	& 0		& & $9/2^+$\\  
 752.2(5) 		& 0.5(1)    & {0.5(1) }
                                            & 1.4(12)	& 3184 	& 2432 	& $(5/2^+)$	    & 	        &2087.7(5) 		& 5.4(5) 	& 8.1(9) 	& 2088	& 0		&   			& $9/2^+$\\  
 768.9(5) 		& 38.4(19)  & 38.5(19) 
                                            & 43.8(40)	& 1220 	& 451	& $(5/2^-)$ 	& $1/2^-$	&2126.9(5) 		& 1.6(3) 	& 1.4(7) 	& 3888	& 1761	&   			& \\ 
 840.1(5) 		& 5.5(5)    & {5.5(5) }
                                            & 10.8(9) 	& 2058 	& 1220 	& 			    & $1/2^-$	&2142.8(6) 		& 1.7(3) 	& 1.4(7) 	& 2143	& 0		& 	    		& $9/2^+$\\ 
 861.8(5) 		& 0.5(1)    & {0.5(1) }
                                            &  			& 2216 	& 1354 	&    			& $13/2^+$	&2155.3(5) 		& {6.8(6)} 	
                                                                                                                                & 6.4(8)	& 3150	& 995	& $(13/2^-)$    & $11/2^+$\\ 
 863.0(5) 		& 4.3(5)    & {4.3(5) }
                                            & 3.8(4) 	& 2551 	& 1688	&    			& $17/2^-$	&2216.4(5) 		& 7.4(6) 	& 7.5(9) 	& 2216	& 0		&   			& $9/2^+$\\ 
 873.1(5) 		& 3.7(4)    & {3.7(4) }
                                            & 3.6(5) 	& 3150 	& 2277 	& $(13/2^-)$	&       	&2268.3(6) 		& 0.7(2) 	& 		    & 3888	& 1620	&   			& \\ 
 891.1(5) 		& 1.0(4)    & {1.0(4)}
                                            & 1.4(12) 	& 2447 	& 1554 	& 	    		& 			&2295.4(5)$^*$ 	& 1.9(3) 	& 1.3(4) 	& 		& 		&   			& \\ 
 895.3(5) 		& 1.2(3)    & {1.2(3)}
                                            & 1.4(12) 	& 2589 	& 1693 	& 		    	& 			&2352.3(6) 		& 0.7(2) 	& 		    &(3347) & 995	& 	            & $11/2^+$\\ 
 914.8(5) 		& 0.6(1)    & {0.6(1) }
                                            & 1.4(12) 	& 3347 	& 2432 	& $(5/2^+)$	    & 			&2356.3(5) 		& 1.1(3) 	& 1.5(7) 	& 4118	& 1761	&   			& \\ 
 922.9(5) 		& 2.5(5)    & {2.5(5) }
                                            &  			& 2277 	& 1354 	& 		    	& $13/2^+$	&2387.6(6)$^*$ 	& 0.8(2) 	&  		    & 		& 		&   			& \\ 
 927.7(5)    	& 1.0(4)    & {1.0(4) }
                                            &  		    & 3150	&(2222) & $(13/2^-)$  	&       	&2415.4(5) 		& 1.2(2) 	& 		    & 3971	& 1554	&   			& \\ 
 967.1(5) 		& 5.0(5)    & {5.0(5) }
                                            & 6.2(7) 	& 3184 	& 2216 	& $(5/2^+)$	    & 			&2431.6(5) 		& 1.2(3) 	& 5.3(6) 	& 2432	& 0		&   			& $9/2^+$\\ 
 994.9(5) 		& 100.0(33) & 100.1(33)  	
                                            & 100.0(51)	& 995 	& 0 	& $11/2^+$	    & $9/2^+$	&2460.4(5) 		& {4.0(4)} 	
                                                                                                                                & 3.8(8) 	& 4082	& 1620	&    		    & \\ 
 1020.1(5)      & 10.3(7) 	&               & 10.1(8)	& 2015 	& 995 	& 			    & $11/2^+$	&2497.8(5)$^*$ 	& 0.7(2) 	& 		    & 		& 		& 			    & \\ 
 1041.0(5) 	    & 1.1(7)  	&               & 2.3(5) 	& 3184  & 2143  & $(5/2^+)$	    &       	&2510.0(6) 		& 0.3(1) 	& 		    & 2510	& 0		& 		    	& $9/2^+$\\ 
 1051.4(5) 	    & 0.7(1)    &               &           & 2134  & 1082  &               & $(3/2^-)$	&2526.3(5) 		& 1.0(3) 	& 		    & 4082	& 1554	& 	    		& \\
 1065.1(5)      & 7.9(6) 	&           	& 8.5(7) 	& 3150 	& 2085 	& $(13/2^-)$    & 	        &2541.3(6) 		& 1.4(2) 	& 		    & 2541	& 0		&    			& $9/2^+$\\ 
 1095.9(5) 	    & 2.9(6) 	&           	& 6.7(7) 	& 3184 	& 2088 	& $(5/2^+)$	    &       	&2605.8(6) 		& 0.7(3) 	& 		    & 2606	& 0		& 			    & $9/2^+$\\ 
 1103.7(5) 	    & 3.6(5) 	&           	& {2.5(13)}       
                                                        & 1554	& 451	&               & $1/2^-$	&2619.4(10)		& 1.5(3)	& 		    & 3701	& 1082	& 		    	& $(3/2^-)$\\ 
 1123.9(5) 	    & 2.4(5) 	&           	& 2.5(13) 	& 3184	& 2058	& $(5/2^+)$	    &       	&2668.8(10)		& 1.4(4) 	& 		    & 3888	& 1220	& 	    		& $(5/2^-)$\\ 
 1130.6(5) 	    & 2.7(7) 	&           	& 3.4(9) 	& 3347 	& 2216 	& $(5/2^+)$	    &           &2861.3(10)		& 1.8(3)	& 		    & 4082	& 1220	&   			& $(5/2^-)$\\
 1134.9(5) 	    & 7.2(10) 	&           	& 6.8(7) 	& 3150 	& 2015 	& $(13/2^-)$    & 			&2880.3(10)$^*$	& 2.5(4) 	& 4.3(10) 	& 		& 		& 			    & \\ 
 1170.1(5)$^*$	& 1.4(5) 	&           	& 			&  		& 		& 			    & 			&2918.1(10)		& 0.9(4) 	& 		    & 3913	& 995	& 			    & $11/2^+$\\ 
 1204.3(5) 	    & 0.5(3) 	&           	& 1.4(12) 	& 3347	& 2143	& $(5/2^+)$	    & 			&2970.7(10)		& 0.3(1) 	& 		    & 3966	& 995	& 			    & $11/2^+$\\ 
 1221.4(5) 	    & 2.0(3) 	&           	& 6.3(9) 	& 2216	& 995	& 			    & $11/2^+$	&2998.7(10)		& 0.5(3) 	& 		    & 4082	& 1082	& 			    & $(3/2^-)$\\ 
 1226.4(5) 	    & 1.0(2) 	&           	& 3.6(27) 	& 2447	& 1220 	& 			    & $(5/2^-)$	&3023.5(12)$^*$	& 0.3(1) 	& 		    & 		& 		& 			    & \\ 
 1227.4(5) 	    & 3.9(17) 	&           	&           &(2222) & 995   & 			    & $11/2^+$	&3150.1(15)		& 0.7(3)	& 		    & 3150	& 0		& $(13/2^-)$	& $9/2^+$\\ 
 1234.4(5) 	    & 6.1(6) 	&              	& 7.7(11) 	& 2589	& 1354	& 			    & $13/2^+$  &3184.0(10)		& {3.9(5)} 	
                                                                                                                                & 4.0(7) 	& 3184	& 0		& $(5/2^+)$	    & $9/2^+$\\ 
 1259.4(5) 	    & 1.0(3) 	&           	& 1.4(5) 	& 3347	& 2088	& $(5/2^+)$	    &       	&3285.2(11)		& 0.9(3) 	& 1.3(4) 	& 3285	& 0		& 			    & $9/2^+$\\ 
 1287.3(5) 	    & 6.9(6) 	&               & 7.0(7) 	& 3347	& 2058	& $(5/2^+)$	    & 			&3347.2(10)		& {1.5(3)} 	
                                                                                                                                & 3.0(10) 	& 3347	& 0		& $(5/2^+)$ 	& $9/2^+$\\ 
 1354.2(5)		& 15.8(10) 	&               & 20.5(12)	& 1354	& 0		& $13/2^+$	    & $9/2^+$	&3384.5(11)$^*$	& 0.6(5)	& 		    & 		& 		& 			    & \\ 
 1363.3(5)		& 1.3(4) 	&           	& 1.0(5) 	& 2447	& 1082	& 			    & $(3/2^-)$	&3486.9(10)		& {2.2(4)} 	
                                                                                                                                & 1.4(6) 	& 3487	& 0		& 			    & $9/2^+$\\ 
 1387.0(5)$^*$ 	& 0.7(3) 	&           	&  			& 		&	 	& 			    & 	        &3701.0(10)		& {6.2(6)}	
                                                                                                                                & 4.7(17)	& 3701	& 0		& 			    & $9/2^+$\\ 
 1396.7(6)$^*$ 	& 3.2(10) 	&           	& 3.0(15) 	& 		& 		& 			    &       	&3887.8(10)		& {1.1(3)} 	
                                                                                                                                & 1.1(7) 	& 3888	& 0		& 			    & $9/2^+$\\ 
 1423.0(5) 	    & 17.7(11)	&              	& 16.9(10)	& 3184	& 1761	& $(5/2^+)$	    & 			&3913.1(10)		& {3.3(4)} 	
                                                                                                                                & 2.7(5) 	& 3913	& 0		& 			    & $9/2^+$\\ 
 1458.0(5) 	    & 1.4(6) 	&           	& 1.4({12}) 		
                                                        & 3150	& 1693	& $(13/2^-)$	& 			&3965.7(10)		& 4.4(5) 	& 5.0(12) 	& 3966	& 0		& 			    & $9/2^+$\\ 
 1462.6(5) 	    & 8.6(7) 	&               & 12.2(12)	& 3150	& 1688	& $(13/2^-)$	& $17/2^-$	&3976.5(10)$^*$ & 0.7(2) 	& 		    & 		& 		& 			    & \\ 
 1500.3(5)$^*$ 	& 3.6(5) 	&           	& 4.4(5) 	& 		& 		& 			    & 			&4117.9(10)		& {2.0(3)} 	
                                                                                                                                & 1.7(5) 	& 4118	& 0		& 			    & $9/2^+$\\ 

\end{tabular}
\end{ruledtabular}
\end{table*}
\endgroup

\subsection{\label{subsec:isomerHalfLife}Half-life of the $\bf{17/2^-}$ isomer at 1688~keV}
	The presence of the $17/2^-$ isomer was first reported by \citeauthor{isomer} and its half-life was measured to be 8.5(5) $\mu$s \citep{isomer}, which has been adopted in ENSDF \citep{ENSDF}. The half-life has been measured in several other studies to be 11(1) $\mu$s \citep{isomer2,isomer3}, 2.2(3) $\mu$s \citep{isomer4}, and 11.2(2) $\mu$s \citep{isomer5}. Although Refs. \citep{isomer2, isomer3} and \citep{isomer5} reported consistent values, it is worthwhile to cross-check these values with our data.
	
\begin{figure}[!htb]
\includegraphics[width=\linewidth]{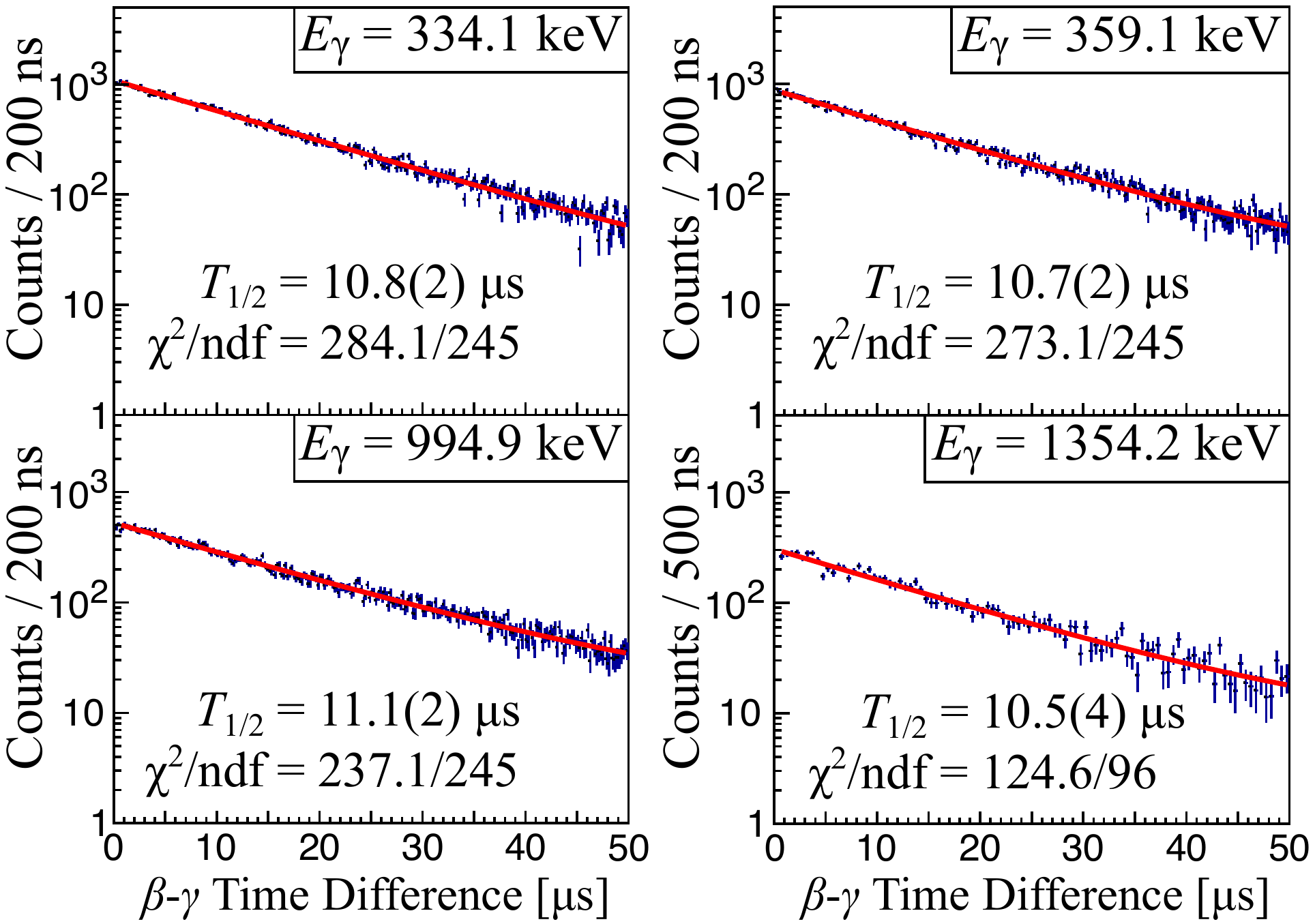}
\caption{\label{fig:isomer4fits}(Color online) The half-life of the $17/2^-$ isomer measured by fitting the time difference of the timestamps of $\beta$-particles and $\gamma$-rays gated on 334.1, 359.1, 994.9, and 1354.2 keV transitions. Both summed time distribution and the weighted average of these four time distribution yields a half-life of 10.8(1) $\mu$s.}
\end{figure}

	In order to determine the half-life of the $17/2^-$ isomer, the time distribution of the number of $\beta$-gated $\gamma$-rays is used. The $\gamma$-rays of interest are 334.1, 359.1, 994.9, and 1354.2 keV following the decay of the $17/2^-$ isomer. The half-life of this isomer is expected to be around 10 $\mu$s, therefore the time window of the event construction was set to be 50 $\mu$s. FIG. \ref{fig:isomer4fits} shows the time distributions of the four respective $\gamma$-rays following the $\beta$-decays. They are constructed by gating on each energy of the $\gamma$-rays with time-random background subtraction on a 2-dimensional histogram that shows the $\beta$-$\gamma$ time difference.
	
	By taking the weighted average of the half-lives obtained from the four $\gamma$-rays, the half-life of the $17/2^{-}$ isomeric state is deduced to be 10.8(1) $\mu$s. A fit performed on the summed time distributions also yields a half-life of 10.8(1) $\mu$s. This value is consistent with the half-life reported in Ref. \citep{isomer2, isomer3}, but not with those reported in Refs. \citep{isomer, isomer4} and slightly below the value reported in Ref. \citep{isomer5}.
	
\section{\label{sec:betaDecay}Properties of the $\beta$-decay of $^{\bf{129}}\mathrm{\bf{Cd}}$}	
The $\beta$-decays of the two states of $^{129}$Cd with $J^{\pi} = 11/2^- \text{ and } 3/2^+$ are studied in the present work. While the relative intensities of the $\gamma$-ray transitions are discussed in the previous section, the $\beta$-feeding intensities from each $\beta$-decaying state in $^{129}$Cd are necessary to determine the characteristics of the $\beta$-decays.

\subsection{\label{subsec:numDecay} Number of $^{129}$Cd decays and absolute $\gamma$-ray intensities}
The first ingredient for determining the $\beta$-feeding intensity is the number of detected $\beta$-decays of $^{129}$Cd. By performing a fit on the summed time distribution of detected $\beta$-particles in the beam cycle (FIG. \ref{fig:cyclefit}, see Section \ref{sec:ExpRes} for the description of the beam cycle), we can deduce the number of detected $\beta$-particles from the decays of $^{129}$Cd and which fraction of the decays populate the $9/2^+$ ground state and the $1/2^-$ $\beta$-decaying isomer in $^{129}$In, respectively.

Five different fixed half-lives have been used: the half-lives of the $1/2^-$ $\beta$-decaying isomer and the $9/2^+$ ground state in $^{129}$In, the two $\beta$-feeding-weighted half-lives of $^{129}$Sn, and an averaged half-life of the two states in $^{129}$Cd (discussed below). The known half-lives of the two $\beta$-decaying states in $^{129}$In are 611(5) ms for the $9/2^+$ ground state and 1.23(3) s for the $1/2^-$ $\beta$-decaying isomeric state \citep{ENSDF}. The two $\beta$-decaying states of the grand-daughter nucleus $^{129}$Sn, which can be populated by the $\beta$-decay of $^{129}$In, are the $3/2^+$ ground state and the $11/2^-$ isomeric state. They have the half-lives of 2.23(4) min and 6.9(1) min, respectively \citep{ENSDF}. In this fit, the half-lives of $^{129}$Sn are weighted according to the population of each state resulting from the $\beta$-decay of the two states in $^{129}$In \citep{decay129In} because both of the $\beta$-decaying states in $^{129}$In can populate the two $\beta$-decaying states in $^{129}$Sn. In this way additional complexity of the fit can be avoided. Nevertheless, the contributions from the decays of states in $^{129}$Sn are very limited due to their significantly longer half-lives compared to $^{129}$Cd or $^{129}$In. 

For $^{129}$Cd, the two $\beta$-decaying states have similar half-lives of $T_{1/2}(11/2^-) = 147(3)$ ms and $T_{1/2}(3/2^+) = 157(8)$ ms \cite{CdHalflife}, therefore decays from both states are fitted with one (averaged) half-life. Since we know which states in $^{129}$In are populated by either of the two $\beta$-decaying states in $^{129}$Cd (see following Section \ref{subsec:betaintensity}), the fraction of decays originating from each of the two states in $^{129}$Cd can be determined from the absolute transition intensities and the $\beta$-feeding intensities to the $9/2^+$ ground state and the $1/2^-$ isomeric state in $^{129}$In. Depending on the composition of these two states in the beam, the averaged half-life can be anywhere between these two half-lives. However, the branching ratios obtained from the fit to the two $\beta$-decaying states in $^{129}$In have to be consistent with the observed number of $\gamma$-rays populating them. As a consequence, the lower limit of this averaged $^{129}$Cd half-life for this fit is obtained to be 153.3 ms, constrained by the number of $\gamma$-rays feeding the $1/2^-$ $\beta$-decaying state in $^{129}$In. The upper limit of the fitted half-life may also be determined in a similar manner, which is deduced to be 164.3 ms. The best fit is obtained when the half-life is fixed to $T_{1/2}(^{129}\mathrm{Cd})=153.6$ ms, as shown in FIG. \ref{fig:cyclefit}. The systematic uncertainty on the number of $\beta$-decays of $^{129}$Cd arising from the uncertainty on the averaged half-life is evaluated by varying its half-life in the fit. 

As a result of this fit, the branching ratios of the $^{129}$Cd decays to the $9/2^+$ ground state and the $1/2^-$ isomer in $^{129}$In are determined to be 68(12)\% 
and 32(12)\%, respectively.
The total number of $\beta$-decays of $^{129}$Cd is determined to be $1.22(2) \times 10^6$. 
From this value and the efficiency-corrected number of 994.9 keV $\gamma$-rays observed, the absolute transition intensity of the strongest transition 994.9 keV, corrected for internal conversion, is determined to be 30(1) per 100 $\beta$-decays. Since the $\gamma$-rays are observed in coincidence with the $\beta$-particles, this absolute intensity can be determined independently of the $\beta$-efficiency of the experimental setup. All the other absolute intensities of $\gamma$-ray transitions may also be determined accordingly, using the relative $\gamma$-ray intensities shown in TABLE \ref{tab:gamma}.

\subsection{\label{subsec:betaintensity} $\beta$-feeding intensities and $\log ft$ values}

  By comparing the sum of absolute intensities of transitions which populate a given excited state to the ones that de-populate this state, the $\beta$-feeding intensities can be calculated. For this calculation, the absolute intensities are corrected for internal conversion (IC). However, in order to determine the $\log ft$ values, it is still necessary to know the fraction of $\beta$-decays originating from the $11/2^-$ and $3/2^+$ states in $^{129}$Cd, respectively.
  
  \citeauthor{EURICA} suggested that nearly all $11/2^-$ decays in $^{129}$Cd proceed via the 994.9 keV and 1354.2 keV transitions or directly populate the $9/2^+$ ground state, based on the information obtained by gating on the transitions from the $(21/2^+)$ isomer to the $11/2^-$ state in $^{129}$Cd \citep{EURICA}. This resulted in the assumption that only the 3150 keV state and the levels fed by its de-excitation are populated by the decay of the $11/2^-$ state in $^{129}$Cd. However, our observation suggests that the decay of the states at 3913 keV and 3966 keV populates the $11/2^+$ state at 994.9 keV in $^{129}$In and it is likely that these two states are also fed from the $\beta$-decay of the $11/2^-$ state. In addition to this, we assume that states at 3971 keV and 4082 keV are populated by the $\beta$-decay of the $3/2^+$ state, since the decay of these two states populates the $1/2^-$ isomer at 451 keV through cascades of transitions. 
  
\begin{figure}
\includegraphics[width=\linewidth]{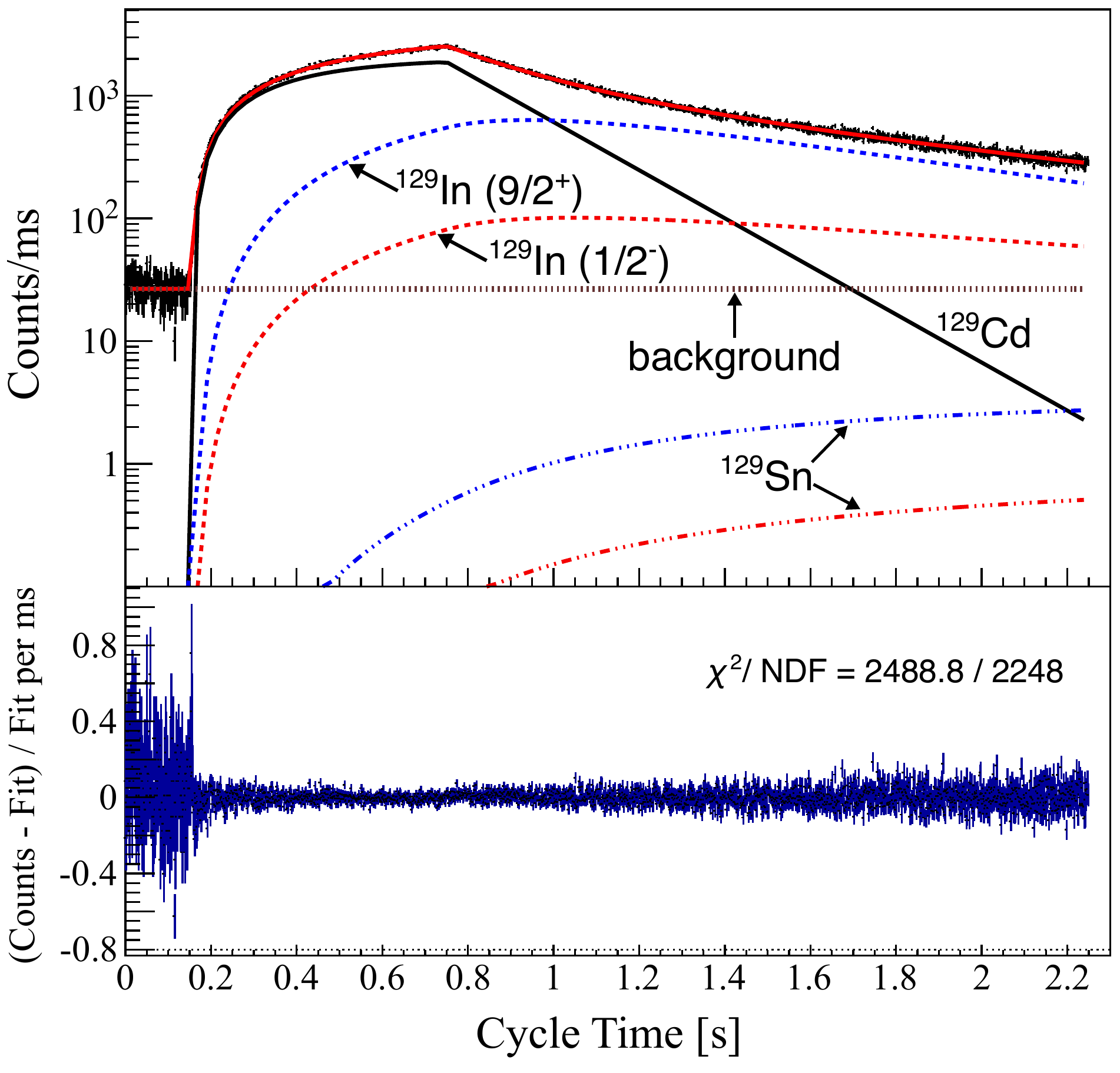}
\caption{\label{fig:cyclefit}(Color online) (Top) Time distribution of detected $\beta$-particles in a beam cycle, simultaneously fitted with the half-lives of $^{129}$Cd, $^{129}$In, and $^{129}$Sn (see text for details). In the figure, the half-life of $^{129}$Cd is set to be 153.6 ms and this half-life results in the best fit. (Bottom) Fit residual distribution.}
\end{figure}

  By adding the absolute intensities of the transitions from the states which are populated by the decays of the $11/2^-$ state in $^{129}$Cd to the ground state, and the $\beta$-feeding intensities to the ground state and the level at 2216 keV, then subtracting the absolute intensities of the 1221.4 keV and 861.8 keV transitions, we conclude that 53(12)\% of all $\beta$-decays of $^{129}$Cd occur from the $11/2^-$ state. Consequently, 47(12)\% of all $\beta$-decays come from the $3/2^+$ state. These $\beta$-feeding intensities to the $9/2^+$ ground state and the $1/2^-$ isomer in $^{129}$In, which are 16(12)\% and 10(12)\%, respectively (TABLE \ref{table:logft}), can be obtained from the IC-corrected number of $\gamma$-rays which populate each state, the branching ratios of the $^{129}$Cd decays to the $9/2^+$ ground state and the $1/2^-$ isomer obtained from the fit shown in FIG.~\ref{fig:cyclefit}, which are 68(12)\% to the $9/2^+$ state and 32(12)\% to the $1/2^-$ state, and the number of observed $^{129}$Cd decays. With this information, we calculate the $\beta$-feeding intensities from each $\beta$-decaying state in $^{129}$Cd. For the calculation of $\log ft$ values, we use the half-lives of $T_{1/2} = 147(3)$ ms for the $11/2^-$ state and $T_{1/2} = 157(8)$ ms for the $3/2^+$ state \citep{CdHalflife}, which were obtained from the time distributions of gated $\gamma$-rays using the same data set as used in this work. The $Q$-value of the $\beta$-decay, $Q_\beta = 9780(17)$ keV, is taken from the latest Atomic Mass Evaluation (AME2016 \citep{AME2016}). 
  
  According to the mass measurement reported in Ref. \citep{ISOLTRAP}, the excitation energy of the $3/2^+$ state in $^{129}$Cd is 343(8)~keV, which makes the $11/2^-$ state the ground state. This is in agreement with the present shell model approach. 
  The results of the calculation using the program \textsc{logft} \citep{logft} are listed in TABLE \ref{table:logft}.
	
\begin{table*}[t]
	\caption{\label{table:logft} $\beta$-decay feeding intensities to each excited state and their $\log{ft}$ values. ``lit" denotes the values from Ref.~\cite{EURICA} for comparison with the current analysis. 
	$I_{\beta^-}(11/2^-)$ and $I_{\beta^-}(3/2^+)$ shows the $\beta$-feeding intensities for the decay of the $11/2^-$ ground state and the $3/2^+$ state at 343(8)~keV in $^{129}$Cd, respectively. 
	The $\log{ft}$ values are calculated based on the $\beta$-feeding intensities for each $\beta$-decaying state in $^{129}$Cd. All limits are quoted as 2$\sigma$ (95\% confidence) limits.
	}
	\begin{ruledtabular}
	\begin{tabular}{l c c c c c c c c c}
	$E_x$[keV] &	$J^{\pi}$  &	$I^{\mathrm{lit}}_{\beta^-}$[\%] &	$I_{\beta^-}$[\%] & $I^{\mathrm{lit}}_{\beta^-}(11/2^-)$[\%] & $I_{\beta^-}(11/2^-)$[\%] & $I^{\mathrm{lit}}_{\beta^-}(3/2^+)$[\%] & $I_{\beta^-}(3/2^+)$[\%] & $\log ft^{\mathrm{lit}}$ & $\log ft$\\
	\hline
	0	    & $9/2^+$   & $<$14     & 16(12)   & $<$28 & 30(23)    &       &                   & $>$5.3 	& $>$4.9    \\
	451	    & $1/2^-$   & $<$9      & 10(12)            &       &           & $<$18 & 22(26)            & $>$5.4 	& $>$4.9    \\
	995	    & $11/2+$   & 6.5(20)   & 5.0(17) 	        & 13(4) & 9.4(38)   &       &                   & 5.4(1) 	& 5.59(20)  \\
	1082	& $(3/2^-)$ & 3.6(6)    & 2.2(7)  	        &       &           & 7(1)  & 4.7(19)           & 5.7(1) 	& 5.96(18)  \\
	1220	& $(5/2^-)$ & 3.0(17)   & 3.3(9)  	        &       &           & 6(4)  & 7.0(26)           & 5.7(6)    & 5.76(19)  \\
	1354	& $13/2^+$  & 3.2(12)   & $<$3.3 	        & 8(3)	& $<$6.3    &       &                   & 5.5(2) 	& $>$5.7  \\
	1554	&           & $<$1.0    & 1.2(4)  	        &       &           & $<$2  & {2.5(10)}           & $>$6.1 	& {6.13(18)}  \\
	1620	&           & 1.2(4)    & 0.9(3)  	        &       &           & 2(1)  & {2.0(8)}            & 6.1(2) 	& {6.21(18)}  \\
	1688	& $17/2^-$  & $<$0.4    & $<$1.3  	        & $<$1  & {$<$2.5}    &       &                   & $>$6.3 	& $>$6.0    \\
	1693	&           & 0.7(6)    & $<$0.9  	        & 1(1)  & {$<$1.7}    &       &                   & 6.3(5) 	& $>$6.2    \\
	1761	&           & 0.5(10)   & {$<$1.2}  	        &       &           & 2(2)  & {$<$2.6}            & 6.0(5) 	& {$>$6.1}    \\
	2015	&           & 1.1(3)    & 0.9(4)  	        & 2(1)  & {1.7(8)}    &       &                   & 5.9(2) 	& {6.08(21)}  \\
	2058	&           & 1.3(6)    & 0.7(3)  	        &       &           & 3(1)  & {1.5(8)}            & 5.8(2) 	& {6.23(24)}  \\
	2085	&           & $<$0.1    & $<$0.2        	& $<$1  & $<$0.5    &       &                   & $>$6.2    & $>$6.6    \\
	2088	&           & $<$0.4    & 1.4(3)  	        &       &           & $<$1  & {3.0(10)}            & $>$6.2 	& {5.92(15)}  \\
	2134	&           &           & 0.6(2)  	        &       &           &       & {1.4(6)}            &           & {6.24(19)}  \\
	2143	&           &           & $<$0.4 	        &       &           &       & {$<$0.8}            &        	& $>$6.5    \\ 
	2216	&           & 1.4(6)    & 0.7(3)  	        & 3(1)  & 1.3(7)    &       &                   & 5.7(2) 	& 6.14(24)  \\
    (2222)  &           &           & 0.9(5)            &       & {1.6(10)}   &       &                   &           & 6.0(3)    \\
	2277	&           &           & {0.6(3)}  	        & $<$1  & {1.2(6)}    &       &                   &           & {6.16(22)}  \\
	2432	&           & $<$1.1    & $<$0.2  	        &       &           & $<$2  & {$<$0.5}            & $>$5.9 	& $>$6.7    \\
	2447	&           & 1.3(11)   & $<$0.3            &       &           & 3(2)  & $<$0.7            & 5.7(3) 	& $>$6.5    \\
	2510	&           &           & 0.4(1) 	        & 	    & {0.7(2)}    &       &                   &        	& {6.31(13)}  \\
	2541	&           &           & 0.6(1) 	        & 	    & 1.1(3)    &       &                   & 	        & 6.13(12)  \\
	2551	&           & $<$0.5    & $<$0.7  	        & $<$1  & {$<$1.3}    &       &                   & $>$6.1    & {$>$6.1}    \\
	2589	&           & $<$1.2    & {$<$0.7}  	        & $<$2  & {$<$1.2}    &       &                   & $>$5.8 	& {$>$6.1}    \\
	2606	&           &           & 0.2(2)  	        & 	    & 0.4(4)    &       &                   &	        & 6.5(5)    \\
	3150	& $(13/2)^-$& 25.3(8)   & {24.1(9)} 	        & 52(5) & {45(10)}    &       &                   & 4.2(1) 	& {4.34(10)}  \\
	3184	& $(5/2)^+$ & 13.3(8)   & {10.4(5)}  	        &       &           & 26(3) & {22(6)}             & 4.5(1) 	& 4.77(12)  \\
	3285	&           & 0.4(1)    & 0.3(1) 	        & 		&           & 1(1)  & {0.6(3)}            & 5.9(5) 	& {6.31(22)}  \\
	3347	& $(5/2)^+$ & 8.6(9)    & {6.1(4)}  	        &       &           & 17(2) & 13(3)             & 4.7(1)    & 4.95(11)  \\
	(3347)  &           &           & 0.2(1)            &       & 0.4(2)    &       &                   &           & {6.33(17)}  \\
	3487	&           & 0.7(2)    & 0.6(1) 	        &       &           & 1(1)  & {1.4(4)}            & 5.8(5) 	& 5.88(13)  \\
	3701	&           & 1.6(6)    & {2.7(2)}  	        &       &           & 3(1)  & {5.8(16)}           & 5.3(2) 	& {5.20(13)}  \\
	3888	&           & 1.2(4)    & {1.5(2)}  	        &       &           & 2(1)  & 3.2(9)            & 5.4(2) 	& {5.40(13)}  \\
	3913	&           & 0.9(2)    & {1.3(2)}        	&       & 3.1(9)    & {2.4(6)}  &                   & 5.4(2) 	& {5.38(11)}  \\
	3966	&           & 1.9(4)    & {1.4(2)}  	        &       & 3.9(10)   & {2.6(7)}  &                   & 5.1(1) 	& {5.33(12)}  \\
	3971	&           & 0.7(1)    & 1.6(2)  	        &       &           &       & {3.5(10)}            &        	& {5.34(13)}  \\
	4082	&           & 2.3(13)   & {2.7(3)}  	        &       &           &       & {5.7(16)}           &        	& {5.09(13)}  \\
	4118	&           & 1.0(3)    & 0.9(1)  	        &       &           & 2(1)  & {2.0(6)}            & 5.3(2) 	& {5.53(14)}  \\
	\end{tabular}
	\end{ruledtabular}
\end{table*}

\section{\label{sec:discussion}Discussion}
\subsection{\label{subsec:logft}$\log ft$ values}
The ($13/2^-$), ($5/2^+$), and other ($5/2^+$) levels at 3150, 3184, and 3347 keV, were previously reported to be populated by $\nu 0 g_{7/2} \rightarrow \pi 0 g_{9/2}$ allowed Gamow-Teller (GT) transitions with the corresponding $\log ft$ values of 4.2(1), 4.5(1), and 4.7(1), respectively \citep{EURICA}. The $\log ft$ values obtained for these states in the current measurement are {4.34(10)}, {4.77}(12), and {4.95}(11), respectively. For the $(13/2^-)$ state at 3150 keV, the $\log ft$ value is in agreement with the previous measurement, and for the other two states the obtained values are slightly higher than the previous measurement. Nonetheless, as discussed in Ref. \citep{EURICA}, it is evident that the decays to these three states show the characteristics of allowed GT transitions. However, the nature of the decays with $\log ft $ values larger than 5.0 remained unclear except for the possibility that first-forbidden (ff) transitions may compete with allowed GT decays. 

\begin{figure}[h]
\includegraphics[width=\linewidth]{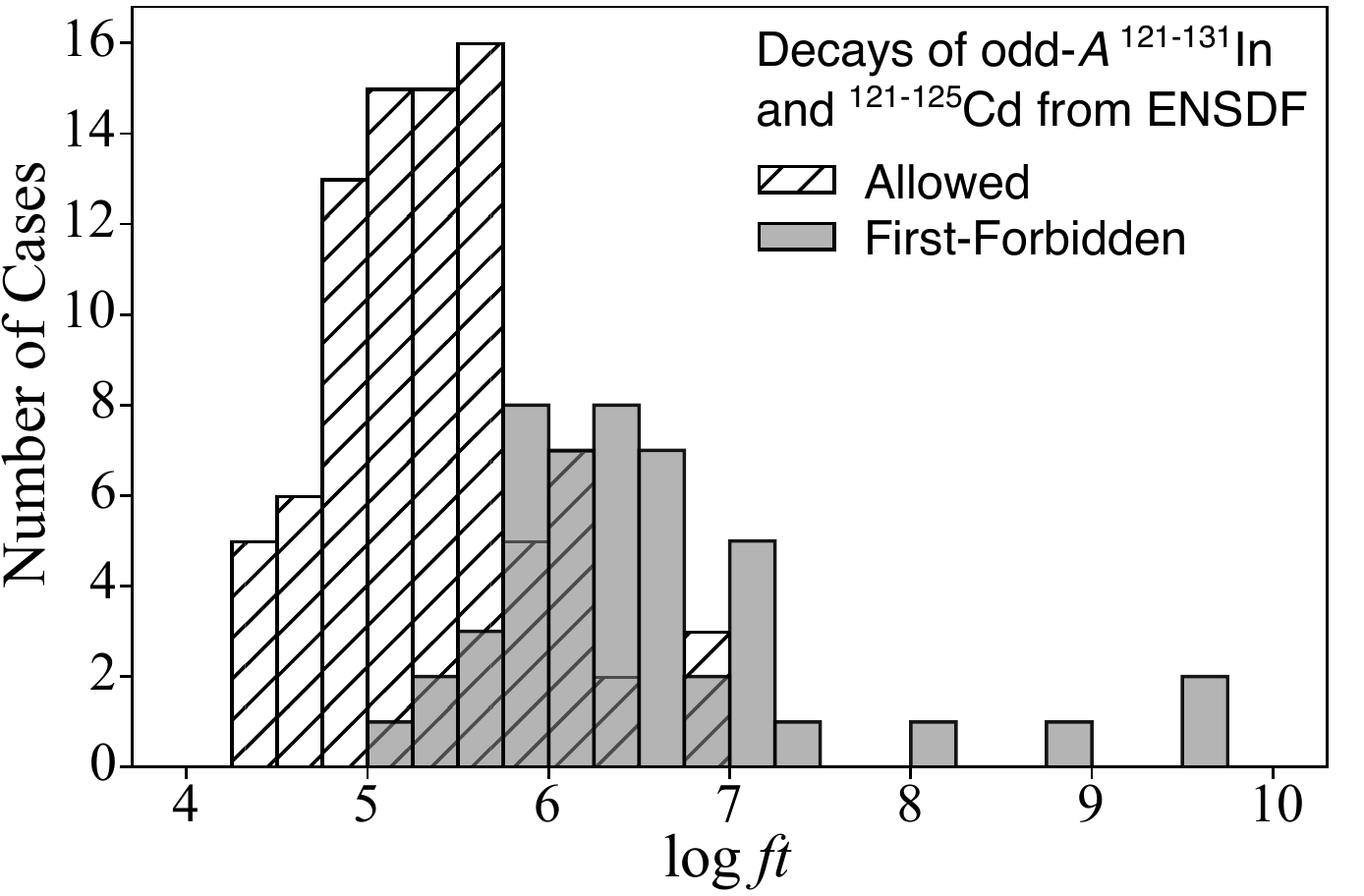}
\caption{{\label{fig:logft}}Distributions of $\log ft$ values for allowed transitions and first-forbidden (ff) transitions in odd-$A$ $^{121\text{-}131}$In and $^{121\text{-}125}$Cd, based on the data from ENSDF \cite{ENSDF}. 
}
\end{figure}

FIG. \ref{fig:logft} shows the distributions of $\log ft$ values for allowed and ff decays in neutron-rich odd-$A$ In and Cd isotopes, based on the data available in ENSDF \cite{ENSDF}. It can be observed that in this region transitions with $\log{ft}$ values $\leq 5.5$ are much more likely to be allowed than ff. Therefore, in this discussion we focus on levels with $\log{ft}\leq 5.5$ as candidates of allowed GT feeding.


Observing the de-excitation patterns of the states above 3.7 MeV, the state at 3701 keV ({$\log ft = 5.20(13)$}) has a decay branch to the $(3/2^-)$ state at 1082 keV, the state at 3888 keV ({$\log ft = 5.40(13)$}) populates the $(5/2^-)$ state at 1220 keV, and the state at 4082 keV ({$\log ft = 5.09(13)$}) decays to both $(3/2^-)$ and $(5/2^-)$ states. We can thus assume that these states are populated by the decay of the $3/2^+$ state in $^{129}$Cd and that the states at 3701, 3888, and 4082 keV are populated by allowed GT decay. Although the states at 3971 and 4118 keV do not have a decay branch to the levels at 1082 or 1220 keV, based on the rather strong $\beta$-feeding and similar $\log{ft}$ values to these three states, it is likely that they also belong to the states populated by allowed GT decay. 

The states at 3913 and 3966 keV with $\log ft$ values of {5.38(11)} and {5.33(12)}, respectively, are likely to be populated by the decay of the $11/2^-$ state in $^{129}$Cd since their observed decay branches are only to the $11/2^+$ state at 995 keV and the $9/2^+$ ground state. If these two states are populated by the allowed GT decay of the $11/2^-$ state in $^{129}$Cd, the observed de-excitation pattern is consistent with the possible range of $J^\pi$ assignment to the states, which are $(9/2^-)$, $(11/2^-)$, and $(13/2^-)$.

\begin{figure}[h]
\includegraphics[width=\linewidth]{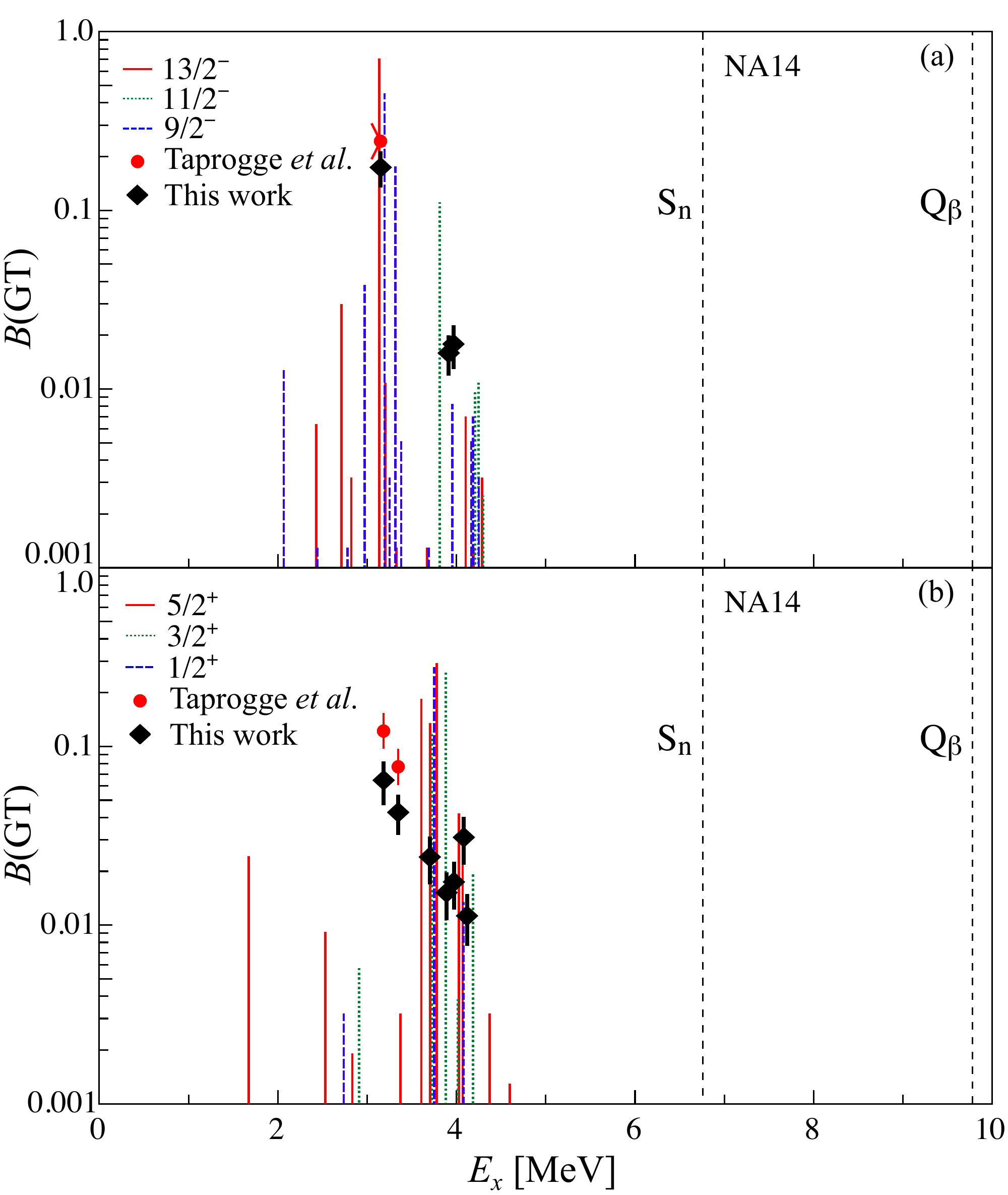}
\caption{{\label{fig:bgt}}(Color online) Observed Gamow-Teller (GT) strength $B(\mathrm{GT})$ in comparison with the calculated distribution for the $11/2^-$ (top) and $3/2^+$ (bottom) states in $^{129}$Cd. The red circles show the results obtained in Ref. \citep{EURICA}. The calculation is identical to the work shown in there.}
\end{figure}

\subsection{\label{subsec:gtstrength} Gamow-Teller strength distribution}
FIG. \ref{fig:bgt} shows the comparison of the distribution of observed and calculated Gamow-Teller strengths, $B(\mathrm{GT})$. The original calculation was carried out in Ref. \cite{EURICA} using the ``NA-14'' interaction derived from the CD-Bonn nucleon-nucleon potential renormalized through the $V_{\text{low-}k}$ approach. The $\pi(f_{5/2},p,g_{9/2}) \nu(g_{7/2},d,s,h_{11/2})$ model space allows to calculate $\nu(g_{7/2}) \rightarrow \pi(g_{9/2})$ GT transitions only. Core excited states populated by $\nu j \rightarrow \pi j$ GT conversion, $j=(g_{7/2},d,s,h_{11/2})$, are excluded in this space. These states are expected to be close to the neutron separation energy of $S_n \approx 6.76$ MeV as the shell gap is $\approx 5.45$ MeV and the $ph$ interaction is repulsive. The corresponding $ph$ configurations, however, may mix into the wave functions of the adopted model space and shift down excitation energies and GT strength. This interaction is shown to be able to reproduce well the low-lying levels in the odd-$A$ $^{125\text{-}129}$Cd and $^{125\text{-}129}$In \cite{cd127}. 

In the current work, the levels at 3913 and 3966 keV are proposed to be populated by the decay of the $11/2^-$ state in $^{129}$Cd through allowed GT transitions. The deduced $B(\mathrm{GT})$ values for these states show a similar distribution as the calculated $B(\mathrm{GT})$ values (FIG. \hyperref[fig:bgt]{\ref{fig:bgt}(a)}). The calculation predicts a stronger fragmentation of the GT strength for the decay of the $3/2^+$ state in $^{129}$Cd which is consistent with the current observation (FIG. \hyperref[fig:bgt]{\ref{fig:bgt}(b)}). 

The levels at 3701, 3888, 3971, 4082, and 4118 keV are proposed to be populated through the allowed GT transition and their $B(\mathrm{GT})$ values agree well with the calculated GT strengths. 

Inspection of the wave functions show that the $I^{\pi}=(9/2\text{ -- }13/2)^-$ daughter states of the $J^{\pi}=11/2^-$ parent are rather purely populated via their $\pi(g_{9/2}) \nu(g_{7/2}h_{11/2})$ 3-hole configuration in the $^{132}$Sn core. The $J^{\pi}=(1/2\text{ -- }5/2)^+$ daughters of the $J^{\pi}=3/2^+$ decay are strongly mixed and populated by their $\pi(g_{9/2}) \nu(g_{7/2}d_{3/2})$ hole content. The latter are specifically sensitive to
mixing of small core-excited components comprising the low-lying $\pi(g_{7/2},d_{5/2})$ orbits, which lie beyond the present model space and may account for the observed upward shift of theoretical strength in FIG. \hyperref[fig:bgt]{\ref{fig:bgt}(b)}. Another consequence of the model space restriction is the overestimate of the GT strengths in spite of the applied quenching factor of 0.75 for the GT matrix element \citep{EURICA}.

In total, 7 states are newly proposed to be populated by the $\nu 0 g_{7/2} \rightarrow \pi 0 g_{9/2}$ allowed GT transition, based on their $\log ft$ values, de-excitation patterns, and comparison with the calculated GT transition strengths. However, due to the large $\beta^-$-decay $Q$-value of 9780(17) keV, it is possible that some high energy $\gamma$-rays are not detected with the current experimental setup, which is known as the "Pandemonium effect" \citep{Pandemonium}. Thus the actual $\beta$-feeding intensities can be smaller than the currently measured values. One has to keep in mind that these possible $J^\pi$ assignments are deduced only from transitions with intensities that are above the detection limit.


\section{Conclusions}
The $\beta$-decay of $^{129}$Cd has been studied utilizing the high efficiency of the GRIFFIN spectrometer. The obtained level scheme of $^{129}$In generally confirms and expands the previously reported results in Ref. \citep{EURICA} with 32 newly observed transitions and 7 newly established levels. The deduced $\beta$-feeding intensities as well as the $\log ft$ values are found to be generally consistent with previously reported values. Based on the distributions of the $\log ft$ values, it is proposed that the 7 observed excited states above 3.7 MeV are fed by the $\nu 0 g_{7/2} \rightarrow \pi 0 g_{9/2}$ allowed GT transitions. This suggests that the allowed GT transition is more dominant than it was previously reported \citep{EURICA}, and the fragmentation of the GT strengths is consistent with theoretical calculations.

\section{Acknowledgements}
The authors would like to thank the operators at the TRIUMF-ISAC facility for providing the radioactive beam. 

This work has been partially supported by the Natural Sciences and Engineering Research Council of Canada (NSERC), the Canada Research Chairs Program, 
NSERC Discovery Grants SAPIN-2014-00028 and RGPAS 462257-2014, the NSERC CREATE Program IsoSiM (Isotopes for Science and Medicine),
the Spanish Ministerio de Econom\'ia y Competitividad under contract No. FPA2017-84756-C4-2-P,
the U.S. National Science Foundation (NSF) under contract NSF-14-01574, 
and the Mexican DGAPA-UNAM program under contract No. PAPIIT-IN110418.

The GRIFFIN spectrometer was funded by the Canada Foundation for Innovation (CFI), TRIUMF, and the University of Guelph. TRIUMF receives federal funding via a contribution agreement with the National Research Council of Canada (NRC). 

Y.S. would like to thank Jorge Agramunt (IFIC Valencia, Spain) for providing a program for fitting the time distribution of $\beta$-particles.

%

\end{document}